\documentclass[11pt,a4paper]{article}
\pdfoutput=1   % required for arXiv submission if pdflatex is used
\usepackage[utf8]{inputenc}
\usepackage{tabularx}
\usepackage{amsmath,amstext,amsfonts,amsbsy,amssymb,amscd,bbm,epsfig,lscape}
\usepackage{color,multirow}

\usepackage{alpha} 
\hypersetup{%
   pdftitle    = {},
   pdfauthor   = {M.Bruno, I.Campos, P.Fritzsch, J.Koponen, C.Pena, D.Preti, A.Ramos, A.Vladikas},
   pdfkeywords = {Lattice QCD, Non-perturbative Effects, Quark masses}
}%
\usepackage{lmodern}
\usepackage[outercaption]{sidecap}
\usepackage{macros}
\usepackage{xspace}
\usepackage{defs}
\usepackage{mathtools}

\newdimen\arrayruleHwidth
\makeatletter
\def\Hline{\noalign{\ifnum0=`}\fi\hrule \@height \arrayruleHwidth
  \futurelet \@tempa\@xhline}
\makeatother
\begin{document}

\preprintno{%
\includegraphics[height=0.8cm]{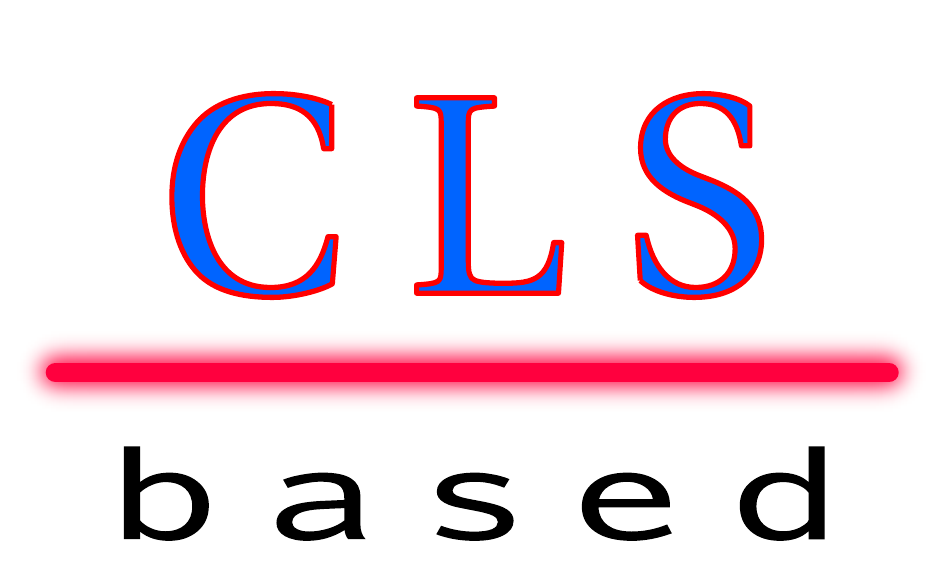}\hfill~\\[-0.8cm]
CERN-TH-2019-174\\
IFT-UAM/CSIC-19-151\\
KEK-CP-372\\
\vfill
}

\title{%
Light quark masses in $\boldsymbol{\Nf=2+1}$ lattice QCD\\
with Wilson fermions
}

\collaboration{\includegraphics[width=2.8cm]{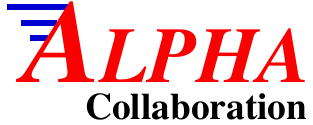}}

\DeclareRobustCommand{\orcid}[1]{\href{https://orcid.org/#1}{\raisebox{-0.2em}{\makebox[1.0em][c]{\includegraphics[width=0.8em]{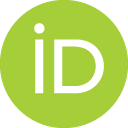}}}}}

\author[cern]{M.~Bruno\orcid{0000-0002-5127-4461}}
\author[ifca]{I.~Campos\orcid{0000-0002-9350-0383}}
\author[cern]{P.~Fritzsch\orcid{0000-0001-8373-3215}}
\author[kek]{J.~Koponen\orcid{0000-0001-9773-5414}}
\author[uam,ift]{C.~Pena\orcid{0000-0003-4830-9245}}
\author[dp]{D.~Preti\orcid{0000-0003-3874-3646}}
\author[tcd]{A.~Ramos\orcid{0000-0003-1654-1816}}
\author[tv]{A.~Vladikas\orcid{0000-0002-4214-0963}}

\address[cern]{Theoretical Physics Department, CERN, CH-1211 Geneva 23, Switzerland}
\address[ifca]{Instituto de F\'{\i}sica de Cantabria IFCA-CSIC, Avda. de los Castros s/n, E-39005 Santander, Spain}
\address[kek]{High Energy Accelerator Research Organization (KEK), Tsukuba, Ibaraki 305-0801, Japan}
\address[uam]{Departamento de F\'{\i}sica Te\'orica, Universidad Aut\'onoma de Madrid, \\Cantoblanco, E-28049 Madrid, Spain}
\address[ift]{Instituto de F\'isica Te\'orica UAM-CSIC, Universidad Aut\'onoma de Madrid, \\C/ Nicol\'as Cabrera 13-15, Cantoblanco, E-28049 Madrid, Spain}
\address[dp]{INFN, Sezione di Torino, Via Pietro Giuria 1, I-10125 Turin, Italy}
\address[tcd]{School of Mathematics \& Hamilton Mathematics Institute, Trinity College Dublin, Dublin 2, Ireland}
\address[tv]{INFN, Sezione di Tor Vergata, c/o Dipartimento di Fisica, Universit\`a di Roma ``Tor Vergata'', \\Via della Ricerca Scientifica 1, I-00133 Rome, Italy}

\begin{abstract}
We present a lattice QCD determination of light quark masses
with three sea-quark flavours ($\Nf=2+1$). Bare quark masses are known
from PCAC relations in the framework of CLS lattice computations with a non-perturbatively
improved Wilson-Clover action and a tree-level Symanzik improved gauge action.
They are fully non-perturbatively improved, including the recently computed Symanzik
counter-term $b_{\rm A} - b_{\rm P}$. The mass renormalisation at hadronic scales
and the renormalisation group running over a wide range of scales are known non-perturbatively
in the Schr\"odinger functional scheme.
In the present paper we perform detailed extrapolations to the physical point,
obtaining (for the four-flavour theory) $\mud(2~{\rm GeV})= 3.54(12)(9)~\MeV$ and
$\mstrange(2~{\rm GeV}) = 95.7(2.5)(2.4)~\MeV$ in the $\msbar$ scheme.
For the mass ratio we have $\mstrange/\mud = 27.0(1.0)(0.4)$.
The RGI values in the three-flavour theory are $\Mud = 4.70(15)(12)~{\rm MeV}$ and $\Ms = 127.0(3.1)(3.2)~{\rm MeV}$.
\\
\end{abstract}


\maketitle

\clearpage

\makeatletter
\g@addto@macro\bfseries{\boldmath}
\makeatother

\section{Introduction}
\label{sec:basics}

The lattice regularisation of QCD provides a well-de\-fined procedure for the
determination of the fundamental parameters of the theory (i.e. the gauge coupling
and the quark masses) from first principles. The aim of the present work is the
determination of the three lightest quark masses (i.e. those of the up, down
and strange flavours), in a framework in which the up and down
quarks are degenerate and all heavier flavours (i.e. charm and above), if present
in the theory,  would be quenched (valence) degrees of freedom. This is known as
$\Nf=2+1$ lattice QCD. Moreover, QED effects are ignored.

The three-flavour theory adopted in this paper is presumably sufficient for 
determining light quark masses due to the decoupling of heavier 
quarks~\cite{Symanzik:1973vg,Appelquist:1974tg,Ovrut:1980uv,Bernreuther:1981sg}.
Indeed, lattice world averages of light quark masses $\mud$, $\mstrange$ do not show
a significant dependence on the number of flavours at low energies for $\Nf\ge 2$ within
present-day errors~\cite{Aoki:2019cca}. This also holds for the more accurately known
 renormalisation group independent ratio $\mud/\mstrange$. Recently, heavy-flavour 
decoupling has been substantiated also non-perturbatively~\cite{Knechtli:2017xgy}. 

This paper is based on large-scale $\Nf=2+1$ flavour ensembles produced by the
Coordinated Lattice Simulation (CLS) effort~\cite{Bruno:2014jqa,Bruno:2016plf}.
The simulations employ a tree-level Symanzik-improved gauge action and a
non-perturbatively improved Wilson fermion action; see
references~\cite{Wilson,Sheikholeslami:1985ij,Luscher:1984xn,Bulava:2013cta}.
The sea quark content is made of a doublet of light degenerate quarks
$m_{\mathrm{q},1} = m_{\mathrm{q},2}$, plus a heavier one $m_{\mathrm{q},3}$.
At the physical point $m_{\mathrm{q},1}, m_{\mathrm{q},2} = 
\mud \equiv\tfrac{1}{2}(\mup+\mdown)$ and $m_{\mathrm{q},3} = \mstrange$.

The bare quark masses produced by CLS~\cite{Bruno:2014jqa,Bruno:2016plf}
need to be combined with renormalisation and improvement coefficients in order
to obtain renormalised quantities with $\rmO(a^2)$ discretisation effects.
We use ALPHA collaboration results for the quark
mass renormalisation and Renormalisation Group (RG) running~\cite{Campos:2018ahf}
in the Schr\"odinger functional (SF) scheme.
Symanzik improvement is implemented for the removal of discretisation effects from correlation functions,
leaving us with $\rmO (a^2)$ uncertainties in the bulk and $\rmO (g_0^4 a)$ ones at the time boundaries. We find that
correlation functions extrapolations to the continuum limit are compatible with an $\rmO (a^2)$ overall behaviour.
The counter-terms
required for the improvement of the axial current are known from
refs.~\cite{Bulava:2015bxa,Bulava:2016ktf,deDivitiis:2017vvw}. The present work has
combined all these elements, obtaining estimates of the up/down and
strange quark masses, as well as their ratio. These are expressed as renormalisation
scheme-independent and scale-independent quantities, known as Renormalisation
Group Invariant (RGI) quark masses. Of course we also give the same results in the
$\msbar$ scheme at scale $\mu = 2$~GeV.

The bare dimensionless parameters of the lattice theory are the strong coupling
$g_0^2 \equiv 6/\beta$ and the quark masses expressed in lattice units 
$a m_{\mathrm{q},1} = a m_{\mathrm{q},2}$ and $a m_{\mathrm{q},3}$, with
$a$ the lattice spacing. They can be varied freely in simulations. Having chosen a 
specific discretisation of the QCD action, its parameters must be calibrated so that 
three ``input'' hadronic quantities (one for each bare mass parameter and one for the lattice spacing)
attain their physical values. Other physical quantities can subsequently be predicted.
Such input quantities are typically very well-known from experiment, but they also need to be
precisely computed on the lattice; examples are ground state hadron masses and 
decay constants $(\mpi,\fpi,\mK,\fK,\ldots)$. Since the majority of numerical large-scale 
simulations do not yet include the small strong-isospin breaking and electromagnetic 
effects, the physical input quantities have to be corrected accordingly. 
Following ref.~\cite{Bruno:2016plf}, we use the values of ref.~\cite{Aoki:2016frl}
\begin{align}\label{eq:input}
 \mpi^\phys &= 134.8(3)\,\MeV \;, &  
 \mK^\phys  &= 494.2(3)\,\MeV \;, \\ \notag
 \fpi^\phys &= 130.4(2)\,\MeV \;, &  
 \fK^\phys  &= 156.2(7)\,\MeV \;. 
\end{align}

The calibration of the lattice spacing, referred to as scale setting,
usually singles out a dimensionful quantity as reference scale 
$\fref\,[\MeV]$. Its dimensionless
counterpart $a\fref$ is computed on the lattice for fixed values of
the bare coupling at the point where the physical spectrum, such as
$[a\mpi/(a\fref)]_{g_0^2}\equiv \mpi/\fref$, is reproduced in the bare parameter
space $(g_0^2,am_{{\rm q},i})$ of the lattice theory. In this way, the lattice
spacings $a(g_0^2)=[a\fref]_{g_0^2}/\fref$, and consequently all computed observables,
are obtained in physical units. 
When simulations approach the point of physical mass parameters
while the lattice spacing is lowered, computational demands rapidly
increase. In the present work results are obtained at 
non-zero lattice spacings and at quark masses which correspond to unphysical
meson and decay constant values.
Thus our data need to be extrapolated to the 
continuum limit and extra/interpolated to the physical quark mass values. This 
is achieved with a joint chiral and
continuum extrapolation.
The present work pays particular attention to these
extrapolations and interpolations and the ensuing sources of systematic error. 

So far we did not specify the reference scale $\fref$. In ref.~\cite{Bruno:2016plf}
the three-flavor symmetric combination 
$\fpiK^{\phys}=\tfrac{2}{3}(\fK^{\phys}+\tfrac{1}{2}\fpi^{\phys})=147.6(5)~\MeV$,
(obtained from the physical input of eqs.~\eqref{eq:input}) was used for calibration, and
for the determination of the hadronic gradient flow scale $t_0$~\cite{Luscher:2010iy}.%
\footnote{The gradient flow scale $t_0$ is defined by the implicit relation $\{t^2 \langle E(t)
          \rangle\}_{t=t_0}=0.3$, for the finite Yang--Mills energy density $E(t)$
          at flow time $t$; see section~6 of ref.~\cite{Bruno:2014jqa}. 
         }%
An artificial (theoretical) hadronic scale with mass dimension -2, $t_0$ is
precisely computable with small systematic
effects~\cite{Sommer:2014mea,Bar:2013ora}, and thus well-suited as intermediate
scale on the lattice. Its physical value determined from CLS
ensembles~\cite{Bruno:2016plf} reads
\begin{equation}\label{eq:t0phys}
        \sqrt{8\tphys} = 0.415(4)(2) \,\fm \;,
\end{equation}
at fixed
\begin{equation}
        \phi_4 \equiv [8t_0 (\mK^2+\tfrac{1}{2}\mpi^2)]^{\phys} = 1.119(21) \;,
\end{equation}
where the first error of $\sqrt{8\tphys}$ is statistical and the second systematic.

The theoretical framework of our work is explained
in Section~\ref{sec:genth}. The definitions
of bare current quark mass\-es, their renormalisation
parameters and the $\rmO(a)$-im\-prove\-ment counter-terms
are provided in standard AL\-PHA-collaboration fashion.
There is also a fairly detailed exposition of how the
so-called ``chiral trajectory" (a Line of Constant Physics --- LCP)
is traced by $\nf=2+1$ CLS simulations. In Section~\ref{sec:simul}
we outline the computations leading to renormalised
current quark masses as functions of the pion squared mass.
These are computed in the SF
renormalisation scheme at a hadronic (low energy) scale.
In Section~\ref{sec:phys-mass} we perform the combined
chiral and continuum limit extrapolations in order to obtain
estimates of the physical up/down and strange quark masses.
Details of the ans\"atze  we have used are provided in
\ref{app-chipt} and in \ref{sec:discretisation}.
Our final results are gathered in Section~\ref{sec:final}.
Preliminary results have been presented in \cite{Bruno:2019xed}.

\section{Theoretical framework}
\label{sec:genth}

We review our strategy for computing light quark mass\-es with improved Wilson fermions.
In what follows equations are often written for a general number of flavours $\nf$.
In practice $\nf = 2+1$. Flavours 1 and 2 indicate the lighter fermion fields, which are degenerate; 
at the physical point their mass is the average up/down quark mass. 
Flavour 3 stands for the heavier fermion, corresponding to the strange quark
at the physical point.

\subsection{Quark masses, renormalisation, and improvement}

The starting point is the definition of bare correlation functions on a lattice with spacing is $a$ and  physical extension
$L^3 \times T$:
\begin{eqnarray}
\fP^{ij}(x_0,y_0) \,\, &\equiv& \,\, - \dfrac{a^6}{L^3} \, \sum_{\vec x, \vec y} \langle P^{ij}(x_0,\vec x) P^{ji}(y_0,\vec y) \rangle \,\, ,
\nonumber \\
\fA^{ij}(x_0,y_0) \,\, &\equiv& \,\, - \dfrac{a^6}{L^3} \, \sum_{\vec x, \vec y} \langle A_0^{ij}(x_0,\vec x) P^{ji}(y_0,\vec y) \rangle \,\, ,
\label{eq:correlPA}
\end{eqnarray}
where the pseudoscalar density and  axial current are
\begin{eqnarray}
P^{ij}(x) \,\, &\equiv& \,\, \bar \psi^i(x) \gamma_5 \psi^j(x) \,\, ,  \\
A_0^{ij}(x) \,\, &\equiv& \,\, \bar \psi^i(x) \gamma_0 \gamma_5 \psi^j(x) \,\, .
\end{eqnarray}
The indices $i,j =1,2,3$ label quark flavours, which are always distinct ($ i \neq j $).

The bare current (or PCAC) quark mass 
is defined via the axial Ward identity at zero momentum and a plateau average between suitable initial and final time-slices $t_{\rm i} < t_{\rm f}$,
\begin{align}\label{eq:mPCAC}
        m_{ij} & \equiv \dfrac{a}{t_{\rm f} - t_{\rm i}+a}\times \\
        & \sum_{x_0 =  t_{\rm i}}^{t_{\rm f}}\dfrac{[\tfrac{1}{2} (\drv0+\drvstar0) \, \fAz^{ij} + \ca a\drv0\drvstar0 \fP^{ij}](x_0,y_0)}{2 \, \fP^{ij}(x_0,y_0)}  \,\, ,\nonumber
\end{align}
with the source $P^{ji}$ positioned either at $y_0=a$ or $y_0=T-a$.\footnote{In our simulations we average correlation functions
with the source at $y_0=a$ and (time-reversed) correlation functions with the source at $y_0=T-a$.
Since bare quantities are computed on lattices with open boundary conditions in time, we do not use translation
invariance for the source position.}
The mass-independent improvement coefficient $\cA$ is determined non-perturbatively~\cite{Bulava:2015bxa}.
The average of two renormalised quark masses is then expressed in terms of the PCAC mass $ m_{ij}$ as follows:
\begin{align}
\label{eq:renmassPCAC}
&\dfrac{m_{i\rm R} + m_{j\rm R}}{2} \,\, \equiv\,\, m_{ij\rm R} \,\, = \dfrac{\ZA(g_0^2)}{\ZP(g_0^2,a\mu)}\,\, m_{ij}\,\, \times \\
& \Big [ 1 \, + \, (\ba-\bp) am_{\mathrm{q},ij} \, + \, (\barbA- \barbP) a \Tr [\Mq] \Big ]
\,\, + \,\, \rmO(a^2) \, ,
\nonumber
\end{align}
where $\Mq \equiv \diag(m_{\mathrm{q},1},m_{\mathrm{q},2}, \cdots , m_{\mathrm{q},\nf})$ is the matrix of the sea quark
subtracted masses, characteristic of Wilson fermions. Given the bare mass parameter $m_{0,i} \equiv (1/\kappa_i - 8)/(2a)$,
with $\kappa_i$ the Wilson hopping parameter, these are defined as 
\begin{equation}
m_{\mathrm{q},i} = 1/(2a\kappa_i) - 1/(2a\kappa_{\rm cr})  \equiv m_{0,i}-\mcr
\end{equation}
where $\mcr \sim 1/a$ is an additive mass renormalisation arising from the loss of chiral symmetry by the regularisation
and $\kappa_{\rm cr}$ is the critical (chiral) point. The average of two subtracted masses is then denoted by
$ m_{\mathrm{q},ij} \equiv \tfrac{1}{2}(m_{\mathrm{q},i}+m_{\mathrm{q},j})$ in eq.~(\ref{eq:renmassPCAC}). 

The axial current normalisation $\ZA(g_0^2)$ is scale-in\-de\-pen\-dent, whereas the current quark 
mass renormalisation parameter $1/\ZP(g_0^2,a\mu)$ depends on the renormalisation scale $\mu$.
The renormalisation condition imposed on the pseudoscalar
density operator $P^{ji}$ defines the renormalisation scheme for the quark masses. The schemes used in the present
work (SF and $\msbar$) are mass-independent. Pertinent details will be discussed in latter sections.

The improvement coefficients $\ba-\bp$ and $\barbA - \barbP$ of eq.~(\ref{eq:renmassPCAC}) cancel $\rmO(a)$ mass-dependent cutoff effects;
they are functions of the bare gauge coupling $g_0^2$. The corresponding counter-terms of eq.~(\ref{eq:renmassPCAC}) contain the subtracted
masses $am_{\mathrm{q},ij}$ and $\Tr [a\Mq]$, which require knowledge on the critical mass $\mcr$. This can be avoided by substituting
these masses with current quark masses and their sum. Their relationship is~\cite{Bhattacharya:2005rb},
\begin{equation}
\label{eq:mij_mqij}
    \mij   = Z  \bigg[ \mqi[ij] + \left( \Zrm - 1 \right) \dfrac{\Tr[\Mq]}{\Nf} \bigg]  + \rmO(a) \,\, ,
\end{equation}
where $Z(g_0^2)\equiv\zp/(\zs \za)$ and $\Zrm(g_0^2)$ are finite normalisations. $\zs$ is the renormalisation
parameter of the non-singlet scalar density $S^{ij} \equiv \bar \psi^i \psi^j$ and $\Zrm/\zs$ is the renormalisation
parameter of the singlet scalar density, which indirectly defines $\Zrm$; cf. ref.~\cite{Bhattacharya:2005rb}.
In the above we neglect $\rmO(a)$ terms, as they only contribute to $\rmO(a^2)$ in the $b$-counter-terms of eq.~(\ref{eq:renmassPCAC}).
Substituting $am_{\mathrm{q},ij}\to am_{ij}$ in the latter expression, we obtain
\begin{align}
\label{eq:renmassPCAC2}
&m_{ij\rm R}(\mu_{\rm had}) \,\, = \,\, \dfrac{\ZA(g_0^2)}{\ZP(g_0^2,a\mu)} \,\, m_{ij} \,\,
\Bigg [ 1 \, + \, (\batil-\bptil) am_{ij} \nonumber \\
&+ \, \Bigg \{ (\batil-\bptil) \dfrac{1-\Zrm}{\Zrm} +
  (\barbA - \barbP) \dfrac{\Nf}{Z \Zrm} \Bigg \} \dfrac{a M_{\rm sum}}{\Nf} \Bigg ] \nonumber \\
&+ \,\, \rmO(a^2) \,\,\, ,
\end{align}
where we define
\begin{eqnarray}
\batil - \bptil & \equiv & \dfrac{\ba - \bp}{Z} \,\, ,
\nonumber \\
M_{\rm sum} & \equiv & m_{12} + m_{23} + \cdots + m_{(\nf-1)\nf} + m_{\nf 1}  \,\, \nonumber \\
& = & \,\, Z \Zrm \Tr[\Mq] + \rmO(a) \,\, . 
\label{eq:Msum}
\end{eqnarray}
To leading order in perturbation theory the difference $\ba-\bp$ is $\rmO(g_0^2)$ and
equals  $\batil-\bptil$. However, non-perturbative estimates
are likely to differ significantly, especially in the range of couplings $g_0$ considered
here ($1.56 \lsim g_0^2 \lsim 1.76$). We will employ non-perturbative estimates
of $\ba-\bp$ and $Z$; cf. ref.~\cite{deDivitiis:2019xla}. The term multiplying
$ M_{\rm sum}$ contains $(1-\Zrm)/\Zrm$ and $(\barbA - \barbP)$. 
In perturbation theory
$\Zrm = 1 + 0.001158\,\CF\,\Nf\,g_0^4$~\cite{Constantinou:2014rka,Bali:2016umi}, 
$(1-\Zrm)/\Zrm \sim  \rmO(g_0^4)$ and $(\barbA - \barbP) \sim \rmO(g_0^4)$~\cite{Bhattacharya:2005rb}.
A first non-perturbative study of the coefficients $\barbA$ and $\barbP$ produced noisy results with 100\% errors~\cite{Korcyl:2016ugy}.
Given the lack of robust non-perturbative results and the fact that the term in curly brackets is $\rmO(g_0^4)$ in perturbation
theory, it will be dropped in what follows.

Once the quark mass averages $m_{12\rm R}$ and $m_{13\rm R}$  are computed say, in 
the SF scheme at a scale $\mu_{\rm had}$, the three renormalised quark masses
can be determined. Since we are working in the isospin limit ($m_{\mathrm{q},1} = m_{\mathrm{q},2}$),
the lighter quark mass is given by $m_{12\rm R}$. Then one can isolate $m_{13\rm R}$ from the ratio
$m_{13\rm R}/m_{12\rm R}$ in which, as seen from eq.~(\ref{eq:renmassPCAC2}), the $ M_{\rm sum}$
counter-term cancels out.

The ALPHA Collaboration is devoting considerable resources to the determination of the non-perturbative evolution of
the renormalised QCD parameters (strong coupling and quark masses) between a hadronic and a perturbative energy scale
($\mu_{\textrm{had}} \leq \mu \leq \mu_{\textrm{pt}}$). Quark mass\-es are renormalised at 
$\mu_{\textrm{had}} \sim \rmO(\Lambda_{\textrm{QCD}})$ and evolved to  
$\mu_{\textrm{pt}} \sim \rmO(M_{\textrm{W}})$~\cite{Luscher:1992zx,Luscher:1993gh,DellaMorte:2004bc,Capitani:1998mq,DellaMorte:2005kg,Fritzsch:2013je,DallaBrida:2016kgh,Bruno:2017gxd,Brida:2016flw,DallaBrida:2018rfy,Campos:2018ahf} in the SF scheme~\cite{Luscher:1992an,Sint:1993un}. Both renormalisation and RG-running are done non-perturbatively. At $\mu_{\textrm{pt}}$ perturbation theory is believed to be reliably controlled
and we may safely switch to the conventionally preferred, albeit inherently perturbative $\msbar$ scheme.

We will be quoting results also for the scheme- and scale-independent renormalisation group invariant (RGI) quark masses $M_{12}$ and $M_{13}$ (corresponding to the current masses $m_{12}$ and $m_{13}$) as well as the physical RGI quark masses $\Mud$ and $\Ms$ derived from them. They
are conventionally defined in massless schemes~\cite{Gasser:1982ap} by
\begin{align}
\label{eq:Mrgi}
M_i \equiv &\m_{i \rm R}(\mu) \left[2b_0 g_{\rm R}^2(\mu)\right]^{-\frac{d_0}{2b_0}} \nonumber \\
& \times \exp\left\{-\int_0^{g_{\rm R}(\mu)}\dif g\left[\dfrac{\tau(g)}{\beta(g)}-\dfrac{d_0}{b_0g}\right] \right\}\,,
\end{align}
for each quark flavour $i$. In our opinion, $M_i$ is better suited for comparisons either to experimental results or other theoretical determinations. 
Equation~\eqref{eq:Mrgi} is formally exact and independent of perturbation theory as long as
the renormalised parameters $(g_{\rm R}, m_{i\rm R})$ and the continuum renormalisation group
functions (i.e. the Callan-Symanzik $\beta$-function and the mass anomalous dimension $\tau$) are known non-perturbatively with satisfacory
accuracy~\cite{Brida:2016flw,DallaBrida:2016kgh,Bruno:2017gxd,DallaBrida:2018rfy,Campos:2018ahf}.
Their computation in the SF scheme with $\Nf=3$ massless quarks has been carried
out in ref.~\cite{Campos:2018ahf}.

Our determination of the renormalised quark masses  is based on the bare current mass averages $m_{ij\rm R}$; cf.
eqs.~(\ref{eq:renmassPCAC}) and (\ref{eq:renmassPCAC2}). 
The analogue of these expressions for the RGI mass averages is given by
\begin{eqnarray}
M_{ij} \equiv \dfrac{1}{2} ( M_i + M_j) = \dfrac{M}{m_{\rm R}(\mu_{\textrm{had}})} \, \, m_{ij\rm R}(\mu_{\rm had}) \,\,\, .
\label{eq:Mrs}
\end{eqnarray}
Note that the ratio $M/m_{\rm R}(\mu_{\textrm{had}})$ is flavour-independent; cf. eq.~(\ref{eq:Mrgi}). In ref.~\cite{Campos:2018ahf} it has been 
computed in the SF scheme for the $\Nf=3$ massless flavours at $\mu_{\textrm{had}} = 233(8)~{\rm MeV}$.

\subsection{The chiral trajectory and scale setting}
\label{subsec:chtraj-scale}

Our aim is to stay on a line of constant Physics within systematic uncertainties of $\rmO(a^2)$,
as we vary the bare parameters of the theory (i.e. the gauge coupling $g_0$ and the  $\nf = 2+1$ quark masses).
In particular, if the improved bare gauge coupling
\begin{equation}
\label{eq:trM0}
\tilde g_0^2 \equiv g_0^2 \,\, \Big ( 1 \, + \, \dfrac{1}{\Nf} b_g(g_0^2) a \Tr [ \Mq ] \Big )
\end{equation}
is kept fixed in the simulations, so does the lattice spacing, with any fluctuations being attributed to  $\rmO(a^2)$-effects.
The problem is that $\bg(g_0^2)$ is only known to one-loop order in perturbation theory~\cite{Sint:1995ch,Sint:1997jx}; $\bg^{\rm\scriptsize PT}=0.012\Nf g_0^2$.
Thus, following refs.~\cite{Bietenholz:2010jr,Bietenholz:2011qq}, we vary the quark masses at fixed $g_0^2$, ensuring that the trace of the quark mass matrix
remains constant:
\begin{equation}
\Tr [ \Mq ] \,\, = \,\, 2 m_{{\rm q},1} \, + \, m_{{\rm q},3} \,\, = \,\, {\rm const} \,\,.
\end{equation}
In this way the improved bare gauge coupling $\tilde g_0^2$ is kept constant at fixed $\beta$ for any  $b_g$.\footnote{In ref.~\cite{Bruno:2016avt} (cf. sect. 5.3.2) it was estimated that when, in some ensembles,  $\Tr [ \Mq ]$ is not constant, the resulting effect on the shift of the lattice spacing is about 6 per mille. This estimate was based on the 1-loop value of $b_g$, $\ba-\bp$ and $Z$.}

\begin{figure*}[t]
   \begin{minipage}{0.375\linewidth}
   \vspace*{1mm}
   \includegraphics[width=\linewidth]{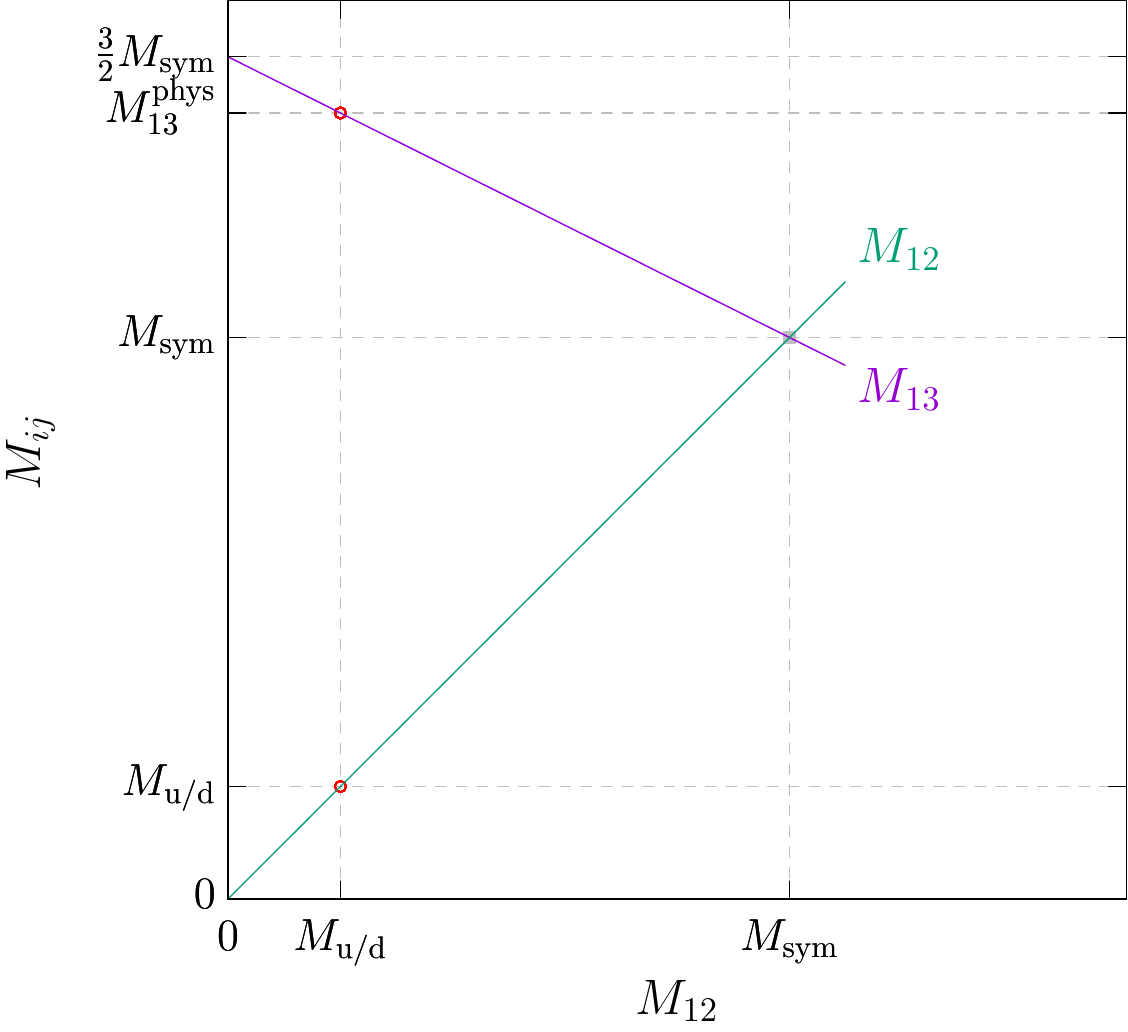}
   \end{minipage}	
   \hspace*{5mm}
   \begin{minipage}{0.565\linewidth}
   \includegraphics[width=\linewidth]{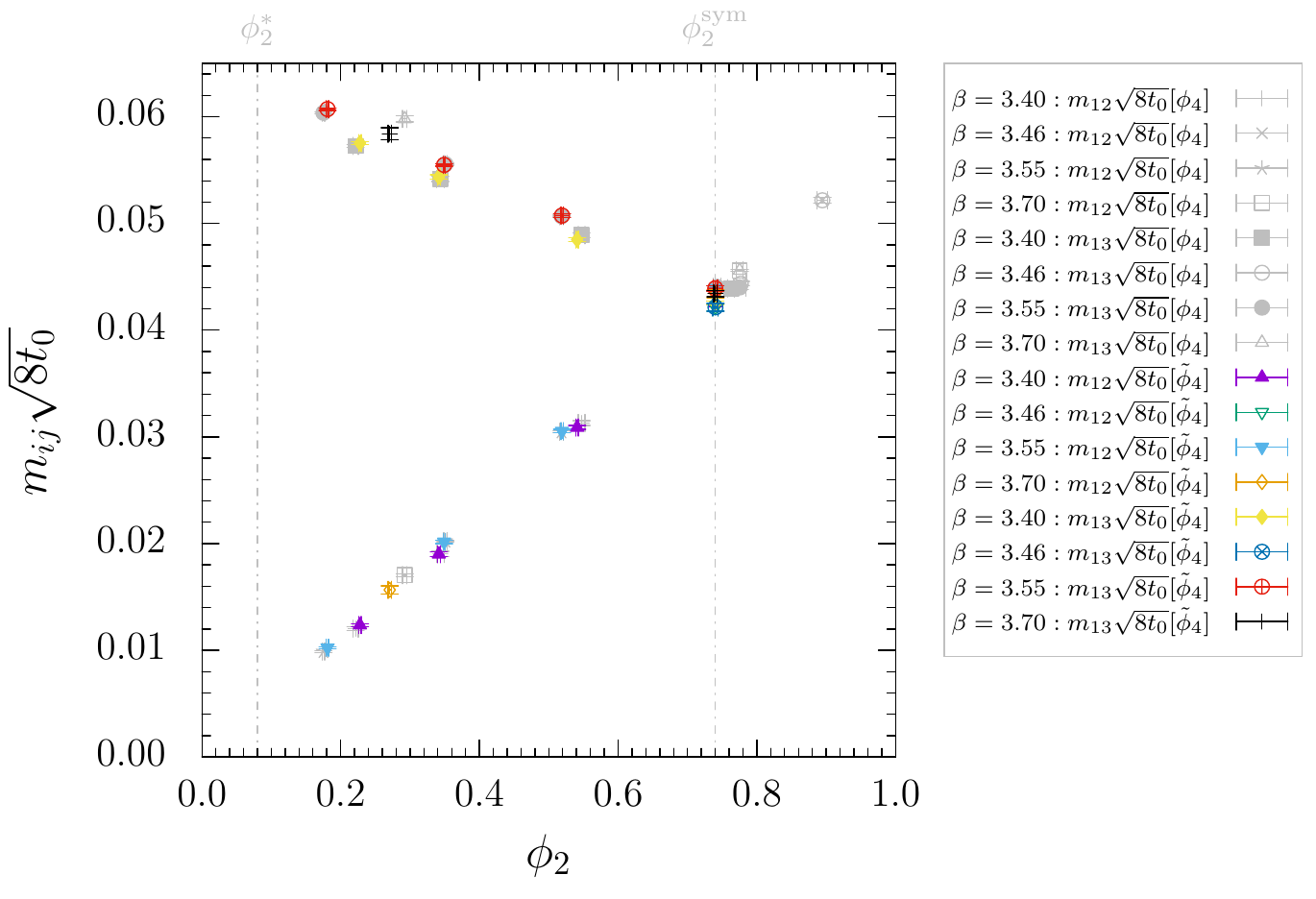}
   \end{minipage}

   \caption{The left panel shows an idealisation of the chiral trajectory for
            renormalised RGI current quark masses $M_{12}$ and $M_{13}$ in the
            continuum. The symmetric point (gray box) is defined by the
            trace of the renormalised RGI quark mass matrix,
            $M_{12}=M_{13}=\Msym=\tfrac{1}{3}\Tr[M]$, and the physical point is
            indicated by red circles, where $\Mphy_{12}=\Mud$ and
            $\Mphy_{13}=\tfrac{1}{2}(\Mud+\Ms)$. The right panel shows our data
            $\phi_{ij}\equiv \sqrt{8t_0} m_{ij}$ versus $\phi_2\equiv 8t_0\mpi^2 \propto m_{12}$.
            Coloured (gray) points correspond to mass-shifted (-unshifted) points in parameter
            space, cf. the discussion in the text.
           }
   \label{fig:Xtraj}
\end{figure*}

This requirement leads to an unusual but unambiguous approach to the physical point, shown in the 
$(M_{12},M_{13})$-plane in the left panel of figure~\ref{fig:Xtraj}. Initially,
one starts at the symmetric point ($am_{\mathrm{q},1}=am_{\mathrm{q},2}=am_{\mathrm{q},3}=a\mqsym$)
for some fixed coupling $\beta=6/g_0^2$, and tunes the mass parameter of
the simulation in such a way that $\Tr[\Mq]=\Tr[\Mq]_{\phys}$ to a good approximation\footnote{In practise one only tunes the bare quark mass $am_{0}$, since $a\mcr$ is unknown a priori, but constant at fixed $\beta$.}.
This is achieved by varying $a\mqsym$ until $(\mK^2+\tfrac{1}{2}\mpi^2)/\fref$
takes its physical value. Since it is proportional to $\Tr[\Mq]$ at leading
order in chiral perturbation theory (\chiPT), it suffices as tuning
observable. In subsequent simulations, one successively lifts the mass-degeneracy
towards the physical point by decreasing the light quark masses while maintaining
the constant-trace condition. By doing so the physical strange quark mass is 
approached from below as in figure~\ref{fig:Xtraj} (left panel). We call this procedure
``the determination of the chiral trajectory''.

Note that the improved renormalised quark mass matrix $M_{\rm R}$ is given by~\cite{Bhattacharya:2005rb}
\begin{align}
  &\Tr [M_{\rm R}] \,\, = Z_m r_m \times\\
  &\Big [ (1 + a \bar d_m \Tr [\Mq] ) \Tr[ \Mq] + a d_m \Tr [\Mq^2] \Big ] +\rmO(a^2) \,\, .\nonumber
\end{align}
Since the $d_m$-counter-term is proportional to squared bare masses, a constant $\Tr [\Mq]$ does not correspond to a constant $\Tr [M_{\rm R}]$; the latter
requirement is violated by $\rmO(a)$ effects. This is an undesirable feature, as it implies that the chiral trajectory is not a line of constant-physics. In practice these violations have been monitored in ref.~\cite{Bruno:2016plf} (Fig.~4, lowest lhs panel), where
$\Tr [M_{\rm R}]$ has been computed, at constant $\Tr [M_q]$, from the current quark masses with 1-loop perturbative Symanzik $b$-coefficients. The violations appear to be bigger than what one would expect from $\rmO(a)$ effects.

These considerations have led the authors of ref.~\cite{Bruno:2016plf} to redefine the chiral trajectory in terms of $\phi_4 =$~const., where
\begin{eqnarray}
\label{eq:defphi4}
\phi_4 \,\, \equiv \,\,  8 \, t_0 \, \Big ( \mK^2 \, + \, \dfrac{1}{2} \mpi^2 \Big ) \,\, ,
\end{eqnarray}
and $t_0 $ is the gluonic quantity of the Wilson flow~\cite{Luscher:2010iy}; it has mass dimension -2. Here
$\mpi$ and $\mK$ are the lightest and strange pseudoscalar mesons respectively; at the physical point these are
the pion $\mpi^\phys$ and kaon $\mK^\phys$.
Keeping $\phi_4$ constant is a Symanzik-improved constant physics condition. But $\phi_4$ is proportional to the sum of the three 
quark masses only in leading-order (LO) chiral perturbation theory ($\chi$PT). Thus, the improved bare coupling
$\tilde g_0^2$ now suffers from $\rmO(a m_q \Tr [M_q ])$
discretisation effects due to higher-order $\chi$PT contributions. In practice, these turn out to be small, as can be
seen from ref.~\cite{Bruno:2016plf} (Fig.~4, lowest rhs panel), where $\Tr [M_{\rm R}]$ has been computed, at constant $\phi_4$. The violations appear to be at most 1\% and thus the variation of the $\rmO(a)$ $b_g$-term in $\tilde g_0^2$ can be ignored.

Obviously, one must also ensure, through careful tuning, that the chosen $\phi_4 =$~const. trajectory passes through the point corresponding to physical up/down and strange renormalised masses (i.e. quark masses that correspond to the physical pseudoscalar mesons $\mpi^\phys$ and $\mK^\phys$). This is done by driving $\phi_4$ to its physical value $\phi_4^\phys = 8 t_0 [ (\mK^\phys)^2 \, + \, (\mpi^\phys)^2/2 ]$ through mass shifts~\cite{Bruno:2016plf}. The aim is to express the computed quantities of interest (in our case the quark masses) as functions of 
\begin{eqnarray}
\phi_2 \,\,  \equiv \,\, 8 \, t_0 \,\mpi^2  \,\, ,
\label{eq:defphi2}
\end{eqnarray}
with $\phi_4$ held fixed at $\phi_4^\phys$, and eventually extrapolate them to $\phi_2^\phys = 8 t_0 (\mpi^\phys)^2$.

The determination of the redefined chiral trajectory is not straightforward. One needs to know the value of $\phi_4^{\rm \phys}$.
The latter is obtained from $t_0$ and the pseudoscalar masses (corrected for isospin-breaking effects)
quoted in eq.~(\ref{eq:input}).
But since the value of $t_0$ is only approximately known, one starts with an initial guess $\tilde t_0$, which provides an initial guess $\tilde \phi_4$.  At each $\beta$, the symmetric point with degenerate masses ($\kappa_1 = \kappa_3$)  is tuned so that the computed $t_0/a^2$, $a \mpi$
and $a \mK$ combine as in eq.~(\ref{eq:defphi4}) to give a  value close to $\tilde \phi_4$.
The other ensembles at the same $\beta$ have been obtained by decreasing the degenerate (lightest) quark mass $m_{{\rm q},1} = m_{{\rm q},2}$, 
while  increasing the heavier mass $m_{{\rm q},3}$ so as to keep $\Tr [\Mq]$ constant. Thus they do not correspond exactly to the same $\tilde \phi_4$. 
Small corrections of the subtracted bare quark masses (or hopping parameters) are introduced, using a Taylor expansion discussed in sect. IV of
ref.~\cite{Bruno:2016plf}, in order to shift $\phi_4$ to the reference value $\tilde \phi_4$ and correct analogously the measured PCAC quark
masses and other quantities of interest such as the decay constants. The procedure is repeated for each $\beta$ and the same
value $\tilde \phi_4$ at the starting symmetric point. 

All shifted quantities are now known at $\tilde \phi_4$ as functions of $\phi_2$. Defining the combination of decay constants
\begin{equation}
\label{eq:fpiK}
f_{\pi{\rm K}} \,\, \equiv \,\, \dfrac{2}{3} \, \Big (f_{\rm K} + \dfrac{f_\pi}{2} \Big ) \,\, ,
\end{equation}
the dimensionless $\sqrt{8 \tilde t_0} f_{\pi{\rm K}}$ is computed for all $\phi_2$ and extrapolated to $\tilde \phi_2 = 8 \tilde t_0 (m_\pi^{\phys})^2$. The extrapolated $\sqrt{\tilde t_0} f_{\pi{\rm K}}$ , combined with the experimentally known $f_{\pi{\rm K}}^{\phys}$, gives a better estimate of $\tilde t_0$, and thus of $\tilde \phi_4$. 
As described  in sect. V  of ref.~\cite{Bruno:2016plf}, this procedure can be recursively repeated and eventually the physical value of $t_0$ is fixed
through $f_{\pi{\rm K}}^{\phys}$; the value in eq.~(\ref{eq:t0phys}) from ref.~\cite{Bruno:2016plf} leads to
\begin{eqnarray}
\phi_4^{\phys} \,\, &=& \,\, 1.119(21) \,\, , \\
\phi_2^{\phys} \,\,  &=& \,\, 0.0804(8)  \,\, .
\label{eq:phi2-phys}
\end{eqnarray}
The main message is that once PCAC quark masses are shifted onto the chiral trajectory defined by the constant $\phi_4^{\phys}$, they only depend on a single 
variable, namely $\phi_2$. 

In analogy to the definitions~(\ref{eq:defphi4}) and (\ref{eq:defphi2}), we also define rescaled dimensionless bare current quark masses and their renormalised counterparts at scale $\mu_{\rm had}$
\begin{equation}
\phi_{ij} \, \equiv \, \sqrt{8 \, t_0} \, m_{ij}  \,\, , \,\,\, \phi_{ij\rm R}(\mu)\, \equiv \, \sqrt{8 \, t_0} \, m_{ij\rm R}(\mu) 
\,\, .
\label{eq:defphiij}
\end{equation}

The redefined chiral trajectory is shown in figure~\ref{fig:Xtraj} (right panel), where
the light-light and heavy-light dimensionless mass averages ($\phi_{12 \rm R}$ and $\phi_{13 \rm R}$ respectively) are plotted as functions of $\phi_2$.  Extrapolating in $\phi_2$ to $\phi_2^{\phys}$ amounts to the simultaneous approach of the light and heavy quark masses to the corresponding physical up/down and strange values. All other physical quantities are then also at the physical point. Sect.~\ref{sec:phys-mass} is dedicated to these extrapolations.

\section{Quark mass computations}
\label{sec:simul}

\begin{table*}
\caption{Details of CLS configuration ensembles, generated as described in ref.~\cite{Bruno:2014jqa}. In the last column, ensembles are labelled by a letter, denoting the lattice geometry, a first digit for the coupling and a further two digits for the quark mass combination.}
\label{tab:CLSens}
\begin{tabular*}{\textwidth}{@{\extracolsep{\fill}}ccccllcccc@{}}
\hline\noalign{\smallskip}
$\beta$& $\dfrac{a}{\fm}$ 
                & $L/a$ & $T/L$ & $\kappa_1$ & $\kappa_3$   & $\dfrac{\mpi}{\MeV}$ 
                                                                    & $\dfrac{\mK}{\MeV}$ 
& $\mpi L$ & label\\
\noalign{\smallskip}\hline\noalign{\smallskip}
 3.40  & 0.086  & 32    & 3     & 0.13675962 & $\kappa_1$   & 420   & 420  & 5.8      & H101 \\
       &        & 32    & 3     & 0.136865   & 0.136549339  & 350   & 440  & 4.9      & H102 \\
       &        & 32    & 3     & 0.136970   & 0.136340790  & 280   & 460  & 3.9      & H105 \\
 &        & 48    & 2     & 0.137030   & 0.136222041  & 220   & 470  & 4.7      & C101 \\ 
\noalign{\smallskip}\hline\noalign{\smallskip}
 3.46  & 0.076  & 32    & 3     & 0.13688848 & $\kappa_1$   & 420   & 420  & 5.2      & H400 \\
\noalign{\smallskip}\hline\noalign{\smallskip}
 3.55  & 0.064  & 32    & 4     & 0.137000   & $\kappa_1$   & 420   & 420  & 4.3      & H200 \\
       &        & 48    & 8/3   & 0.137000   & $\kappa_1$   & 420   & 420  & 6.5      & N202 \\
       &        & 48    & 8/3   & 0.137080   & 0.136840284  & 340   & 440  & 5.4      & N203 \\
       &        & 48    & 8/3   & 0.137140   & 0.136720860  & 280   & 460  & 4.4      & N200 \\
 &        & 64    & 2     & 0.137200   & 0.136601748  & 200   & 480  & 4.2      & D200 \\
\noalign{\smallskip}\hline\noalign{\smallskip}
 3.70  & 0.050  & 48    & 8/3   & 0.137000   & $\kappa_1$   & 420   & 420  & 5.1      & N300 \\
       &        & 64    & 3     & 0.137123   & 0.1367546608 & 260   & 470  & 4.1      & J303 \\
\noalign{\smallskip}\hline
\end{tabular*}
\end{table*}

We base our determination of quark masses on the CLS ensembles for $\Nf = 2+1$ QCD, listed in
Table~\ref{tab:CLSens}. The bare gauge action is the L\"uscher-Weisz one, with tree-level coefficients~\cite{Luscher:1984xn}. 
The bare quark action is the Wilson, Symanzik-improved~\cite{Sheikholeslami:1985ij} one. The Clover term coefficient $\csw$
has been tuned non-perturbatively in ref.~\cite{Bulava:2013cta}.
Boundary conditions are periodic in space and open in time, as detailed in ref.~\cite{Luscher:2012av}.

For details on the generation of these ensembles see ref.~\cite{Bruno:2014jqa}. As seen in Table~\ref{tab:CLSens},  results 
have been obtained at four lattice spacings in the range $0.05 \lsim a/\fm \lsim 0.086$.
For each lattice coupling $\beta = 6/g_0^2$, gauge field ensembles have been generated for a few\footnote{We note in passing 
that for the ensemble with $\beta = 3.46$ we only have results for degenerate quark masses.}
values of the Wilson hopping parameters $\kappa_1 = \kappa_2$ and $\kappa_3$.
The light pseudoscalar meson (pion) varies between 200~MeV and
420~MeV. The heaviest value corresponds to the symmetric point where the three quark masses
and the pseudoscalar mesons are degenerate. The strange meson (kaon) varies between 420~MeV and 470~MeV. Given that our
lightest pseudoscalars are relatively heavy (200 MeV), the chiral limit ought to be taken with care. 

The bare correlation functions $\fP^{ij}, \fA^{ij}$ of eqs.~(\ref{eq:correlPA}) are estimated with stochastic sources located
on time slice $y_0$, with either $y_0 = a$ or $y_0 = T-a$. From them the current quark masses $m_{12}, m_{13}$ are computed
as in eq.~(\ref{eq:mPCAC}), with the $\rmO(a)$-improvement coefficient $c_{\rm A}$ determined non-perturbatively in ref.~\cite{Bulava:2015bxa}.
The exact procedure to select the plateaux range in the presence of open boundary conditions has been explained in
refs.~\cite{Bruno:2014jqa,Bruno:2014lra,Bruno:2016avt}.

Having obtained the bare current quark masses $m_{12}$, $m_{13}$ at four values of the coupling $g_0^2$, we construct the renormalised dimensionless quantities $m_{12\rm R}(\mu_{\textrm{had}})$ and $m_{13\rm R}(\mu_{\textrm{had}})$; cf. eq.~(\ref{eq:renmassPCAC2}).
For this we need the ratio $Z_{\rm A}(g_0^2)/Z_{\rm P}(g_0^2,\mu_{\rm had})$ and the Symanzik $b$-counter-terms. Results for the axial current normalisation $Z_{\rm A}(g_0^2)$ are available in ref.~\cite{DallaBrida:2018tpn}, from a separate computation based on the chirally rotated Schr\"odinger Functional setup
of refs.~\cite{Sint:2010eh,Sint:2010xy,Brida:2014zwa}. The computation of $Z_{\rm P}(g_0^2,\mu_{\rm had})$ in the SF scheme,
for $\mu_{\rm had} = 233(8)~{\rm MeV}$, was carried out in ref.~\cite{Campos:2018ahf} for a theory with $\Nf=3$ massless quarks and the lattice action of the present work.
The $Z_{\rm P}$ results, shown in eqs.~(5.2) and (5.3) of ref.~\cite{Campos:2018ahf}, are in a range of inverse gauge couplings which covers the $\beta \in [3.40,3.85]$ interval of the large volume simulations of ref.~\cite{Bruno:2016plf}, from which our bare dimensionless PCAC masses are extracted.

Besides the ratio $Z_{\rm A}/Z_{\rm P}$, we also need the improvement coefficient $(\batil-\bptil)$, which multiplies the $\rmO(a)$ counter-term proportional
to $a m_{ij}$ in eq.~(\ref{eq:renmassPCAC2}).
To leading order in perturbation theory $\batil-\bptil = -0.0012 g_0^2$. Non-perturbative estimates based on a coordinate-space
renormalisation scheme have been provided for $\nf=2+1$ lattice QCD in ref.~\cite{Korcyl:2016ugy}. More accurate non-per\-tur\-ba\-tive results have been subsequently obtained by the ALPHA Collaboration, using suitable combinations of valence current quark masses, measured on ensembles with $\nf=3$ nearly-chiral sea quark masses in small physical volumes~\cite{deDivitiis:2017vvw,deDivitiis:2019xla}. Also these simulations have been carried out in an inverse coupling range that
spans the interval $\beta \in [3.40,3.85]$ of the large volume CLS results of ref.~\cite{Bruno:2016plf}. They are expressed in the form of ratios $R_{\rm AP}$ 
and $R_{\rm Z}$, from which $(\bA - \bP)$ and $Z$ are estimated; thus $(\batil-\bptil) = R_{\rm AP}/R_{\rm Z}$. In ref.~\cite{deDivitiis:2019xla}, results are
quoted for two values of constant Physics, dubbed LCP-0 and LPC-1. In LCP-0, $R_{\rm AP}$ and $R_{\rm Z}$ are obtained with all masses in the chiral limit. In LCP-1, one valence flavour is in the chiral limit (so it is equal to the sea quark mass), while a second one is held fixed to a non-zero value. The physical volumes are always kept fixed. In ref.~\cite{deDivitiis:2019xla}, eqs.~(5.1), (5.2) and (5.3) refer to LCP-0 results, while those in eqs.~(5.1), (5.4) and (5.5) refer to LCP-1; differences are due to $\rmO(a)$ discretisation effects.

We have opted to use the LCP-0 values of $\batil-\bptil$ in the present work. The covariance matrices of the fit parameters of $R_{\rm AP}$ as well as those of $R_{\rm Z}$ are provided in ref.~\cite{deDivitiis:2019xla}. We {\it assume} that the covariance matrix between fit parameters of $R_{\rm AP}$ and $R_{\rm Z}$
is nil. This is justified {\it a posteriori}, by repeating the analysis with LCP-1 values, as a means to estimate the magnitude of systematic errors arising from our choice. Moreover, we have also compared our LCP-0 results to those obtained from different fit functions, used in the preliminary analysis of ref.~\cite{deDivitiis:2017vvw}, as well as from the perturbative estimate $\batil-\bptil$. We find that the contribution arising from such variations is below $\sim 1\%$ of the total error on renormalised quark masses at the physical point.

As discussed in Section~\ref{sec:genth}, the complicated Symanzik counter-term in curly brackets, multiplying $aM_{\rm sum}$ in eq.~(\ref{eq:renmassPCAC2}), is $\rmO(g_0^4 a)$ in perturbation theory. As there are no robust non-perturbative estimates of its magnitude at present, we will
drop this term, {\it assuming} that the $\rmO(g_0^4 a)$ effects it would remove are subdominant compared to $\rmO(a^2)$ uncertainties.

\begin{table*}
 \caption{Rescaled dimensionless current quark masses $\phi_{12}$ and $\phi_{13}$,
 renormalised in the SF scheme at $\mu_{\rm\scriptscriptstyle had}$, for each CLS ensemble
 used in our analysis.
 Note that for simulation points H102, H105, C101 more than one independent ensembles exist,
 which have been run with different algorithmic setups; we keep those separate before fits.
 All points have been shifted to the target chiral trajectory as
 described in the text, and the quoted errors contain both statistical uncertainties
 and the contribution from renormalisation and the mass shift.
 }
 \label{tab:latticedata}
\begin{tabular*}{\textwidth}{@{\extracolsep{\fill}}lllllll@{}}
\hline\noalign{\smallskip}
\multicolumn{1}{c}{$\beta$} & \multicolumn{1}{c}{ensemble} & \multicolumn{1}{c}{$t_0/a^2$} & \multicolumn{1}{c}{$\phi_2$} & \multicolumn{1}{c}{$\phi_{12}$} & \multicolumn{1}{c}{$\phi_{13}$} & \multicolumn{1}{c}{$\phi_{12}/\phi_{13}$} \\
\noalign{\smallskip}\hline\noalign{\smallskip}
3.40 & H101     & 2.857(13) & 0.747(18)  & 0.0917(26) & 0.0917(26) & 1          \\
     & H102r001 & 2.877(19) & 0.547(20)  & 0.0673(27) & 0.1047(28) & 0.643(10)  \\
     & H102r002 & 2.883(18) & 0.549(19)  & 0.0667(28) & 0.1039(29) & 0.642(10)  \\
     & H105     & 2.886(11) & 0.346(20)  & 0.0416(25) & 0.1167(27) & 0.357(15)  \\
     & H105r005 & 2.896(38) & 0.355(20)  & 0.0420(33) & 0.1160(34) & 0.362(19)  \\
     & C101     & 2.900(19) & 0.238(24)  & 0.0279(31) & 0.1236(27) & 0.226(21)  \\
     & C101r014 & 2.899(14) & 0.233(20)  & 0.0273(26) & 0.1223(30) & 0.222(17)  \\
\noalign{\smallskip}\hline\noalign{\smallskip}
3.46 & H400     & 3.656(20) & 0.747(18)  & 0.0923(28) & 0.0923(28) & 1          \\
\noalign{\smallskip}\hline\noalign{\smallskip}
3.55 & N202     & 5.161(23) & 0.747(18)  & 0.0978(26) & 0.0978(26) & 1          \\
     & N203     & 5.138(16) & 0.526(19)  & 0.0684(27) & 0.1128(26) & 0.606(10)  \\
     & N200     & 5.155(16) & 0.356(18)  & 0.0455(25) & 0.1232(25) & 0.369(13)  \\
     & D200     & 5.171(16) & 0.189(20)  & 0.0237(27) & 0.1232(29) & 0.176(17)  \\
\noalign{\smallskip}\hline\noalign{\smallskip}
3.70 & N300r002 & 8.592(41) & 0.747(18)  & 0.0988(29) & 0.0988(30) & 1          \\
     & J303     & 8.628(40) & 0.278(20)  & 0.0364(31) & 0.1326(39) & 0.274(19)  \\
\noalign{\smallskip}\hline
\end{tabular*}
\end{table*}

As already explained in subsect.~\ref{subsec:chtraj-scale}, our analysis is based on the rescaled dimensionless quantities defined in eqs.~(\ref{eq:defphi4}), (\ref{eq:defphi2}), and (\ref{eq:defphiij}). At each $\beta$ value and for each gauge field configuration, we have results for $t_0/a^2$, $a m_{12}$ and $a m_{13}$ from refs.~\cite{Bruno:2014jqa,Bruno:2016plf},
from which $\phi_{12}$ and $\phi_{13}$ are obtained.
The error analysis is carried out using
the Gamma method approach~\cite{Wolff:2003sm,Schaefer:2010hu,virotta:phdthesis,dobs:alpha} and automatic differentiation for error propagation,
using the library described in ref.~\cite{Ramos:2018vgu}.
This takes into account all the existing errors and correlations in the data and ancillary
quantities (renormalisation constants, improvement coefficients, etc.), and estimates
autocorrelation functions (including exponential tails) to rescale the uncertainties
correspondingly. Following~\cite{Bruno:2016plf}, the estimate of the exponential
autocorrelation times $\tau_{\rm\scriptscriptstyle exp}$
used in the analysis is the one quoted in~\cite{Bruno:2014jqa}, viz., 
\begin{gather}
\label{eq:tauexp}
\tau_{\rm\scriptscriptstyle exp} = 14(3)\,\dfrac{t_0}{a^2}\,.
\end{gather}
We have checked that without attaching exponential tails
statistical errors are 40 to 70\% smaller in our final results.
The full analysis has been crosschecked by an independent code based on (appropriately) binned jackknife error estimation.
Note that one of the strengths of data analysis based on the
Gamma-method is that each Monte Carlo ensemble is treated independently, and the
final statistical uncertainty is determined as a sum in quadratures of the
statistical fluctuations for each ensemble. This allows to trace back which fraction of the statistical variance comes from each ensemble or ancillary quantities, such as renormalisation constants (see references 5-7 in \cite{Ramos:2018vgu}). This feature will be exploited in the error budgets provided below.

The starting values for $\phi_{12\rm R}$ and $\phi_{13 \rm R}$ on which the analysis is based are shown
in Table~\ref{tab:latticedata}, where renormalised quark masses are in the SF scheme at a scale
$\mu_{\rm\scriptscriptstyle had} = 233(8)~{\rm MeV}$.
By suitably fitting these quantities as functions of $\phi_2$, and extrapolating to $\phi_2^{\phys}$, we obtain the results for physical up/down and strange quarks at scale $\mu_{\rm had}$, as detailed in sect.~\ref{sec:phys-mass}. Only then do we convert them to the RGI masses, by multiplying them with the RG-running factor~\cite{Campos:2018ahf}
\begin{equation}
\label{eq:ZMovermR}
\dfrac{M}{m_{\rm R}(\mu_{\textrm{had}})}=0.9148(88) \,\, ,
\end{equation}
with the error added in quadrature; cf.~eq.~(\ref{eq:Mrs}).

Before presenting our chiral fits in section~\ref{sec:phys-mass}, we conclude this section with a
comment on finite-volume effects. Current quark masses are not expected to be affected by
finite-volume corrections, since their values are fixed by Ward identities. On the other hand, meson
masses, decay constants, and the ratio $t_0/a^2$ are expected to suffer from such effects.
This can be directly checked in the ensembles H200 and N202, obtained at $\beta = 3.55$ with degenerate masses and corresponding to volumes of about 2~fm and 3~fm respectively. A glance at the relevant entries of Table II of ref.~\cite{Bruno:2016plf} shows that quark masses do not change as the volume is varied, while meson masses and decay constants vary by about 2.5\%, which corresponds to differences of about $2-3.5\sigma$.\footnote{The ensemble H200 is only used in this context in the present work. Since at $\beta = 3.55$ we have results at two larger volumes (N202/203/200 and D200), we do not use H200 results in our analysis.}
Standard ${\rm SU}(3)$ $\chi$PT NLO formulae are available for masses and decay constants~\cite{Colangelo:2005gd};
$t_0/a^2$ does not suffer from finite-volume effects up to NNLO corrections~\cite{Bar:2013ora}.
In particular, the $\chi$PT-predicted effects for 
meson masses are below the percent level, since the lattice spatial size in units of the inverse
lightest pseudoscalar meson mass is in the range $[3.9,5.8]$.
On the other hand, by directly comparing the values in Table~\ref{tab:latticedata} obtained
at the same lattice spacing and sea quark masses but different volumes
(cf. ref.~\cite{Bruno:2016avt}), it is seen that the finite-volume effects on
$t_0/a^2$ and $m_\pi^2$ are comparable and come with opposite signs. As a result,
they largely cancel in $\phi_2$, the variable in which chiral fits are performed.
Decay constants, which generally suffer from larger finite-volume effects than meson masses,
enter our computation indirectly only --- firstly
through NLO terms in chiral fits, where the finite-volume correction is sub-leading,
and secondly through the physical value of $\sqrt{8t_0}$ determined in~\cite{Bruno:2016plf},
where these corrections have already been taken into account.
We therefore expect that the quantities
most affected by finite-volume effects are the rescaled current quark masses $\phi_{12},\phi_{13}$,
due to the presence of $\sqrt{8t_0}/a$ in their definition. As mentioned above, these are much smaller than our
statistical uncertainty, cf.~Table~\ref{tab:latticedata}.
In the rest of our analysis we will therefore neglect this source of uncertainty.

\section{Extrapolations to physical quark masses}
\label{sec:phys-mass}

Having obtained the dimensionless renormalised current mass combinations $\phi_{12 \rm R}$ and $\phi_{13 \rm R}$ at each
$\beta$ as functions of $\phi_2$, we now proceed with the determination of the physical values $\phi_{\rm ud}$
and $\phi_{\rm s}$. This is done in fairly standard fashion through fits and extrapolations. To begin with,
we note that the two lighter degenerate quark masses are simply given by $\phi_{12}$, whereas the heavier strange 
one is obtained from the difference\footnote{Henceforth all quark masses will be renormalised. In order to simplify the notation, 
we shall drop the subscript R from $\phi_{12 \rm R}$, $\phi_{13 \rm R}$ in this section, in \ref{app-chipt}
 and in \ref{sec:discretisation}.}
\begin{equation}
\phi_{\rm h} \,\, = \,\, 2 \, \phi_{\rm 13} \,- \, \phi_{\rm 12} \,\, .
\end{equation}
It is then possible to perform simultaneous fits of $\phi_{12}$ and $\phi_{\rm h}$ as functions of
$\phi_2$ and the lattice spacing, subsequently extrapolating the results to $\phi_2^{\phys}$ of eq.~(\ref{eq:phi2-phys})
and the continuum limit, so as to obtain $\phi_{\rm ud}$ and $\phi_{\rm s}$. Variants of this method consist in simultaneous fits and extrapolations of either
$\phi_{\rm 13}$ or $\phi_{\rm 12}$ on one hand and their ratio $\phi_{\rm 12}/\phi_{\rm 13}$
on the other. These turn out to be advantageous, as does a certain combination of ratios involving $\phi_{12}$, $\phi_{13}$, $\phi_2$, and $\phi_4$,
for reasons discussed below. We recall in passing that in the ratio  $\phi_{12}/\phi_{13}$ all renormalisation factors cancel.

We use fits based in chiral perturbation theory ($\chi$PT fits) which are expected to model the data well close to the chiral limit $\phi_2 = 0$. Recall that we have performed $\Nf = 2+1$ simulations on a chiral trajectory; starting from a symmetric point where all quark masses are degenerate,
we increase the mass of the heavy quark while decreasing that of the light one, until the physical point is reached. Since both masses are varying,
it is natural to use $\rm SU(3)_L \otimes SU(3)_R$ chiral perturbation theory, which bears explicit dependence on both masses. This works when
{\it all} three quark masses in the simulations are light enough for say, NLO $\chi$PT with three flavours to provide reliable fits. In ref.~\cite{Allton:2008pn}
it is stated that this is the case for their data, obtained with domain wall fermions, as long as the average quark mass satisfies $a m_{\rm avg} < 0.01$.
As seen in Table~2 of ref.~\cite{Bruno:2016plf}, our PCAC dimensionless quark masses $a m_{12}$ and $a m_{13}$ also satisfy this empirical constraint. 
The real test comes about {\it a posteriori}, when the $\rm SU(3)_L \otimes SU(3)_R$ NLO ans\"atze are seen to fit our results well. 

In \ref{app-chipt} and~\ref{sec:discretisation}, ans\"atze for NLO $\chi$PT and discretisation effects are adapted to our specific parametrisation in terms of $\phi_2$ and $\phi_4$. For the current quark masses these are
\begin{align}
\label{eq:phi12_chipt}
\phi_{12} & = \phi_2\left[p_1 + p_2\phi_2 + p_3 K\left(\mathcal{L}_2 - \dfrac{1}{3}\mathcal{L}_\eta\right)\right] \nonumber \\
&+ \dfrac{a^2}{8t_0}\left[C_0+C_1\phi_2\right]\,,\\
\label{eq:phi13_chipt}
\phi_{13} & = \phi_K\left[p_1 + p_2\phi_K + \dfrac{2}{3}p_3 K\mathcal{L}_\eta\right]\nonumber \\
&+ \dfrac{a^2}{8t_0}\left[\widetilde C_0+ \widetilde C_1\phi_2\right]\,,
\end{align}
where $\phi_K = (2 \phi_4 - \phi_2)/2$. The constants $p_1, p_2, p_3$ and $K$ are related to standard $\chi$PT parameters in eqs.~(\ref{eq:p1})-(\ref{eq:K}), whereas the chiral logarithms $\mathcal{L}_2$ and $\mathcal{L}_\eta$ are defined in eq.~(\ref{chilog-2}). For justification of the ansatz used for the discretisation effects, see comments after 
eqs.~(\ref{eq:fll}) and (\ref{eq:flh}). We stress again that  $\phi_{12}$ and $\phi_{13}$ are functions of $\phi_2$ only, $\phi_4$ being held constant. They have common fit parameters $p_1$, $p_2$ and $p_3$. 

Using the above expressions and consistently neglecting higher orders in the continuum $\chi$PT terms, we obtain the ratio of PCAC masses (cf. eqs.~(\ref{eq:phirat_chipt}) and~(\ref{eq:ratio-12-13-discr}))
\begin{align}
\label{eq:phirat_chipt_disc}
\dfrac{\phi_{12}}{\phi_{13}} &= \dfrac{2 \phi_2}{2 \phi_4 - \phi_2}\left[1 + \dfrac{p_2}{p_1} \left(\dfrac{3}{2}\phi_2 - \phi_4\right) - \tilde K\left(\mathcal{L}_2-\mathcal{L}_\eta\right)\right]\nonumber \\
&+ \dfrac{a^2}{8t_0} (2\phi_4 - 3\phi_2) \Big [ D_0 + D_1\phi_2 \Big ] \,\, .
\end{align}
As discussed in \ref{sec:discretisation}, the form of the cutoff effects respects the constraint $\phi_{12}/\phi_{13}=1$ at the symmetric point $m_{{\rm q},1} = m_{{\rm q},3}$, which is exact at all lattice spacings by construction.

For the combination defined in eq.~(\ref{eq:golden_chipt}), we have
\begin{align}
\label{eq:golden_chipt_disc}
\dfrac{4 \phi_{13}}{2\phi_4 - \phi_2} + \dfrac{\phi_{12}}{\phi_2} &= 3p_1 + 2p_2\phi_4 + p_3 K\left(\mathcal{L}_2+\mathcal{L}_\eta\right)\nonumber \\
&+ \dfrac{a^2}{8t_0} \Big [ G_0 + G_1\phi_2 \Big ] \,\,\, .
\end{align}
An alternative to NLO $\chi$PT fits is the use of power series, based simply on Taylor expansions around the symmetric point $m_{{\rm q},1} = m_{{\rm q},2} = m_{{\rm q},3}$,
for which $\phi_2^{\rm\scriptscriptstyle sym} = 2\phi_4^{\rm\scriptscriptstyle phys}/3$:
\begin{eqnarray}
\label{eq:taylor12}
\phi_{12} & = & s_0 + s_1 (\phi_2-\phi_2^{\rm\scriptscriptstyle sym}) + s_2 (\phi_2-\phi_2^{\rm\scriptscriptstyle sym})^2 \nonumber \\
&+& \dfrac{a^2}{t_0}\left[S_0 + S_1(\phi_2-\phi_2^{\rm\scriptscriptstyle sym})\right] \,\, ,\\
\label{eq:taylor13}
\phi_{13} & = & s_0 + \tilde s_1(\phi_2-\phi_2^{\rm\scriptscriptstyle sym}) + \tilde s_2 (\phi_2-\phi_2^{\rm\scriptscriptstyle sym})^2 \nonumber \\
&+& \dfrac{a^2}{t_0}\left[S_0 + \tilde S_1 (\phi_2-\phi_2^{\rm\scriptscriptstyle sym})\right] \,\, .
\end{eqnarray}
Note that imposing the constraint $\phi_{12}=\phi_{13}$ at the symmetric point implies that $s_0$ and $S_0$ are common fit parameters. These expansions are expected to give reliable results in the higher end of the $\phi_2$ range, underperforming close to the chiral limit. They are thus complementary to the chiral fits, which 
are better suited for the small-mass regime. In this sense the two approaches may provide a handle to estimate the systematic uncertainties due to these fits and extrapolations.

We explore various fit variants, in order to unravel the presence of potentially significant 
systematic effects. They are encoded as follows:
\begin{itemize}
\item \underline{Fitted quantities and ans\"atze}:
\begin{itemize}
\item[{\bf\texttt{[chi12]}}] Fit of $\phi_{12}$ data only, using the $\chi$PT ansatz.
\item[{\bf\texttt{[chi13]}}] Fit of $\phi_{13}$ data only, using the $\chi$PT ansatz.
\item[{\bf\texttt{[tay12]}}] Fit of $\phi_{12}$ data only, using the Taylor expansion ansatz.
\item[{\bf\texttt{[tay13]}}] Fit of  $\phi_{13}$ data only, using the Taylor expansion ansatz.
\item[{\bf\texttt{[chipc]}}] Combined fit to $\phi_{12}$ and $\phi_{13}$, using $\chi$PT.
\item[{\bf\texttt{[chirc]}}] Combined fit to $\phi_{13}$ and $\phi_{12}/\phi_{13}$, using $\chi$PT.
\item[{\bf\texttt{[chirr]}}] Combined fit to the ratio $\phi_{12}/\phi_{13}$ and the combination $2\phi_{13}/\phi_K+\phi_{12}/\phi_2$ using $\chi$PT.
\item[{\bf\texttt{[tchir]}}] Combined fit to $\phi_{13}$ and the ratio $\phi_{12}/\phi_{13}$, using the Taylor expansion for $\phi_{13}$ and $\chi$PT for $\phi_{12}/\phi_{13}$.
\end{itemize}

\item \underline{Discretisation effects}:
\begin{itemize}
\item[{\bf\texttt{[a1]}}] Fits with terms $\propto a^2/t_0$ only.
\item[{\bf\texttt{[a2]}}] Fits with terms $\propto a^2/t_0$ and $\propto \phi_2 a^2/t_0$.
\end{itemize}

\item \underline{Cuts on pseudoscalar meson masses}:
\begin{itemize}
\item[{\bf\texttt{[420]}}] Fit all available data, including the symmetric point; i.e. data satisfies $m_\pi \lesssim 420~\MeV$.
\item[{\bf\texttt{[360]}}] Fit excluding the symmetric point; i.e. data satisfies $m_\pi \lesssim 360~\MeV$.
\item[{\bf\texttt{[300]}}] Fit only points for which $m_\pi \leq 300~\MeV$.
\end{itemize}

\end{itemize}
Any given fit will thus be labelled as \texttt{[xxxxx][yy][zzz]}, using the above tags.

\begin{figure*}[t!]
  \centering
  \includegraphics[width=0.85\textwidth]{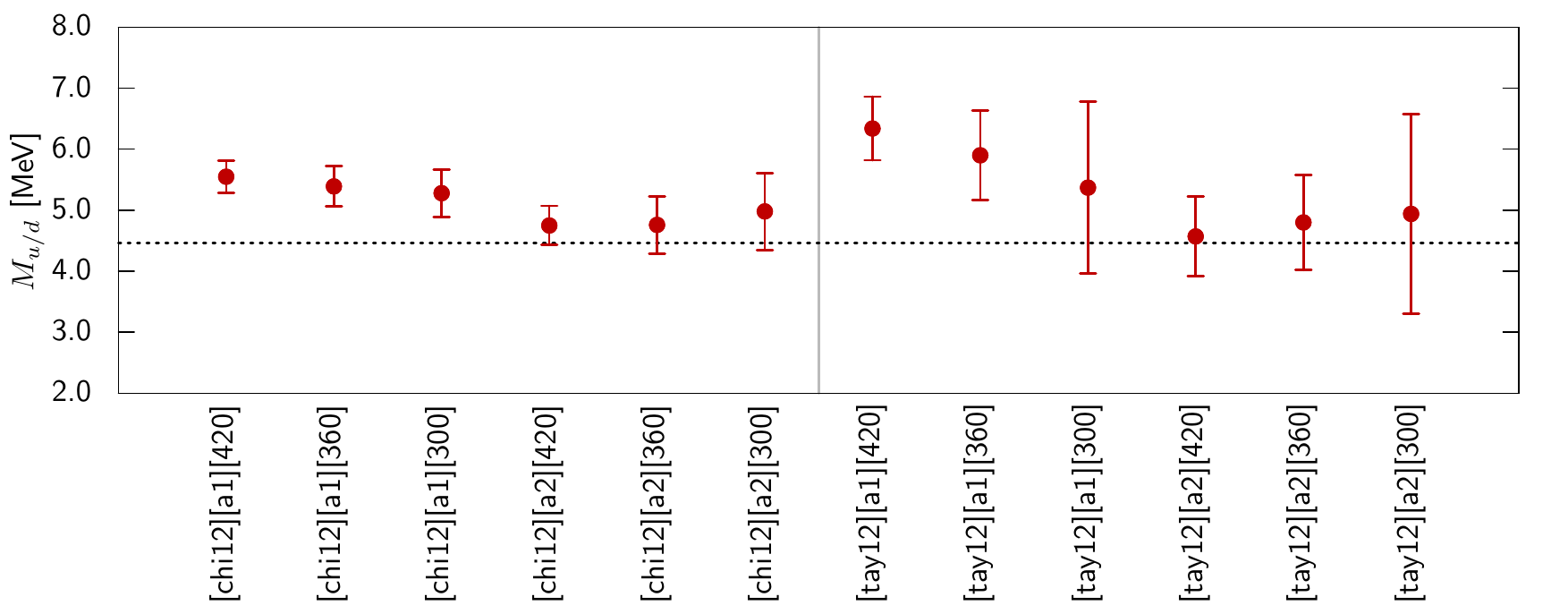}\\[10.0ex]
  \includegraphics[width=0.85\textwidth]{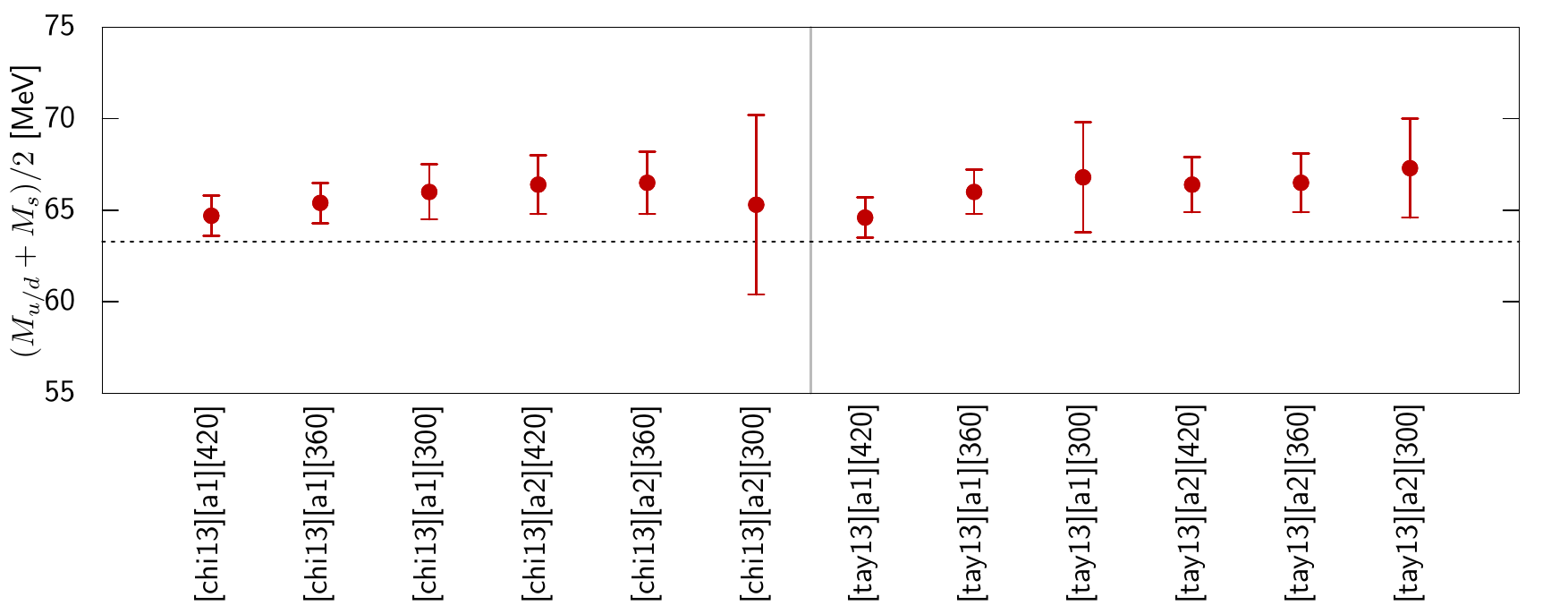}
  \caption{Results for the RGI light ($M_{u/d}$) and averaged ($M_{13}^{\phys}=(M_{u/d}+M_s)/2$) quark masses from
  independent fits to either $M_{12}$ or $M_{13}$. Results are converted to
  $\MeV$ by dividing out with $\sqrt{8t_0^{\rm\scriptscriptstyle phys}}$.
  Dotted lines indicate the central value of the latest FLAG average~\cite{Aoki:2019cca} for reference.}
  \label{fig:masses1}
\end{figure*}

\begin{figure*}[t!]
  \centering
  \includegraphics[width=0.85\textwidth]{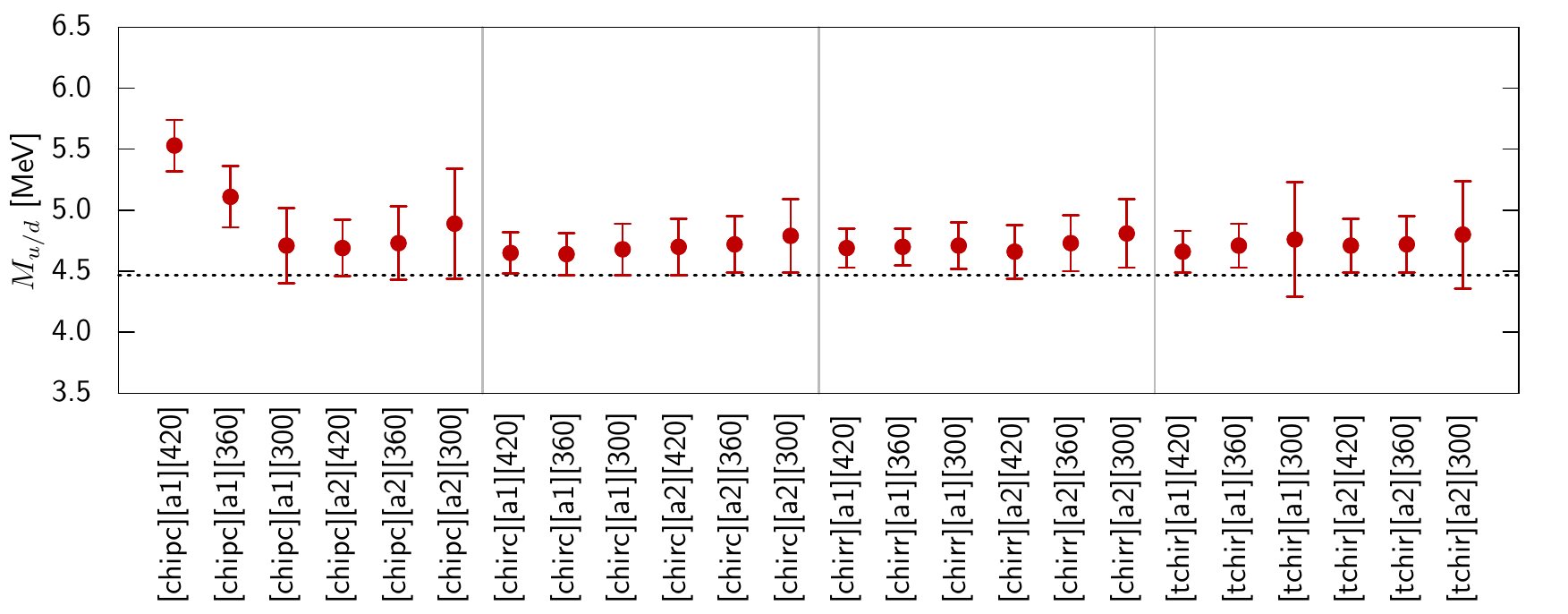}
  \includegraphics[width=0.85\textwidth]{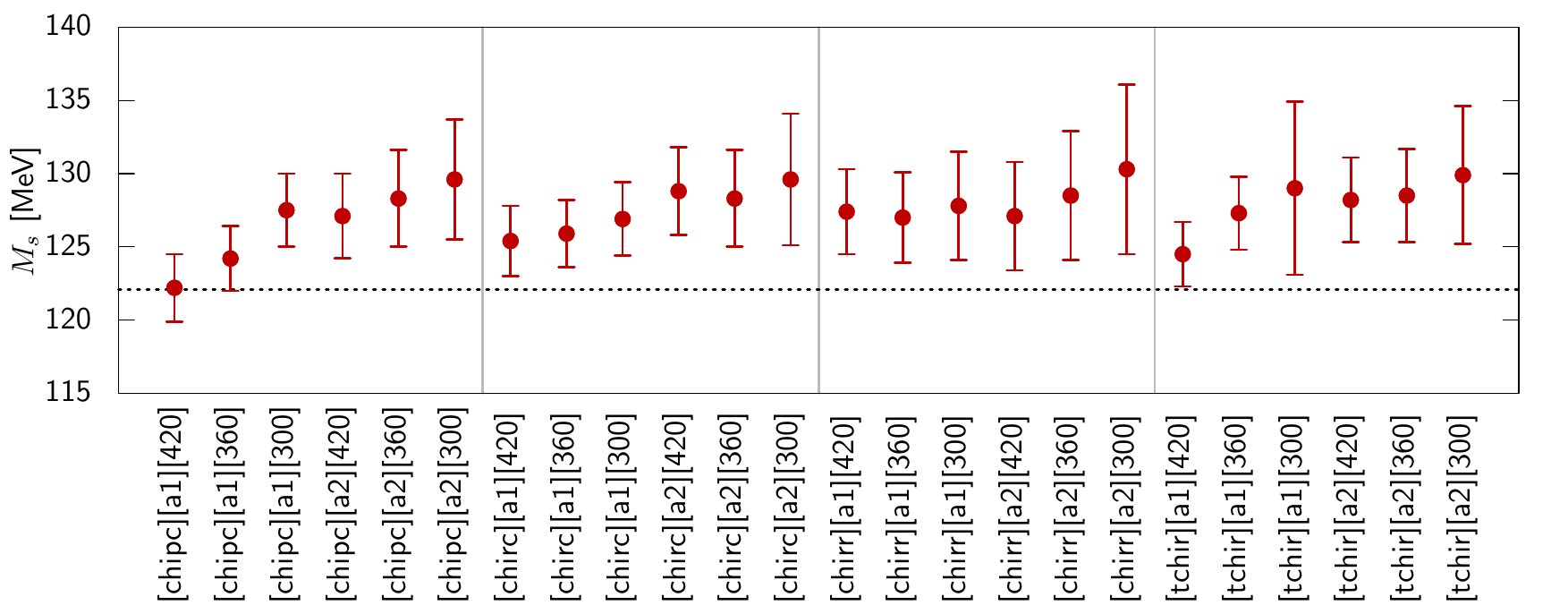}
  \includegraphics[width=0.85\textwidth]{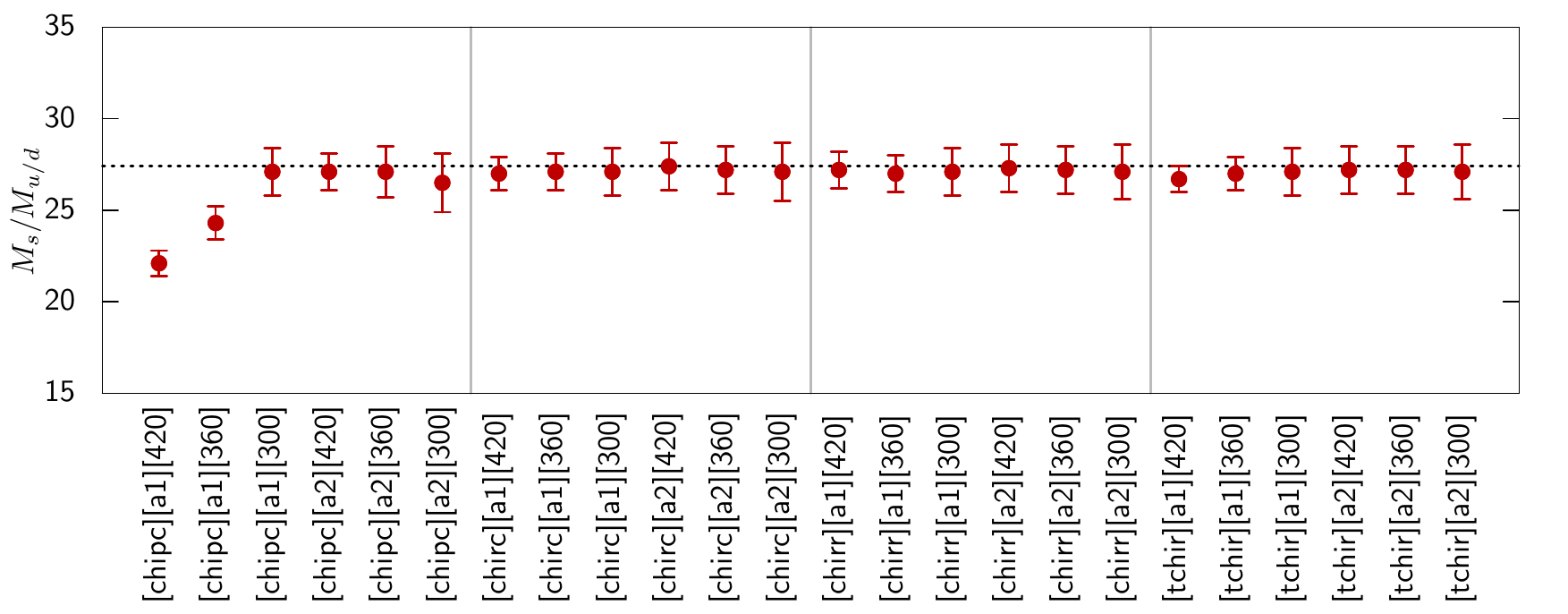}
  \caption{Results for the RGI light ($\Mud$) and strange ($\Ms$) quark masses,
  and their ratio, from
  the simultaneous fits \texttt{[chipc]}, \texttt{[chirc]}, \texttt{[chirr]}, and \texttt{[tchir]}.  Results are converted to
  $\MeV$ by dividing out with $\sqrt{8t_0^{\rm\scriptscriptstyle phys}}$.
  Dotted lines indicate the central value of the latest FLAG average~\cite{Aoki:2019cca} for reference.}
  \label{fig:masses2}
\end{figure*}

The results obtained with the various fit methods at the physical point (cf.~eq.~(\ref{eq:phi2-phys})) are
expressed in physical units by dividing them out by  $\sqrt{8t_0^{\rm\scriptscriptstyle phys}}$.
Multiplication by the factor of eq.~(\ref{eq:ZMovermR}) subsequently gives the RGI mass estimates
shown in Figs.~\ref{fig:masses1} and \ref{fig:masses2}.
We comment on the various fit ans\"atze:

\vspace{5truemm}

\noindent\underline{Independent fits of $\phi_{12}$ and $\phi_{13}$:} comparing light quark masses $\Mud$ (upper panel of 
Fig.~\ref{fig:masses1}) from \texttt{[chi12][a1]} and \texttt{[chi12][a2]} we find that they are sensitive to the presence of
a discretisation term $\propto a^2/t_0$, albeit within $\sim 1-2\sigma$. This difference is attenuated when the 
more stringent mass cutoff \texttt{[chi12][300]} is enforced, mainly because the error increases as less points are fitted. 
The same qualitative conclusions are true for the Taylor expansion fits \texttt{[tay12]} of the light quark mass. 
On the other hand, the lower panel of Fig.~\ref{fig:masses1} shows that the average quark mass $M_{13}^{\phys}$ is not sensitive
to the details of the fit ans\"atze. This is not surprising, given that our simulations have been performed in a region
of rather heavy pions 220~MeV $\leq m_\pi \leq$ 420~MeV, with data covering the physical point $M_{13}^{\phys}$, while $\Mud$
requires long extrapolations. The conclusion is that independent fits are reliable for $\phi_{13}$ but less so for $\phi_{12}$, and
so we discard their results.

\vspace{5truemm}

\noindent\underline{Combined fits to $\phi_{12}$ and $\phi_{13}$:} Fig.~\ref{fig:masses2} shows that the fits
\texttt{[chipc][a1]} and \texttt{[chipc][a2]} give results which are sensitive to the ansatz employed for the cutoff effects. 
This is more pronounced for $\Mud$ and the ratio $\Ms/\Mud$, but persists also for $\Ms$. 
Moreover, fits \texttt{[chipc][a1][420]} and \texttt{[chipc][a1][360]} display visible differences when compared to fits of the
\texttt{[chipc][a2]} variety; the latter agree with results obtained from different fit ans\"atze. For these reason we have
also discarded results from this analysis.

\vspace{5truemm}

\noindent\underline{Combined fits to $\phi_{13}$ and $\phi$-ratios:} As previously ex\-plained, we have explored three ans\"atze, namely
\texttt{[chirc]}, \texttt{[chirr]}, and \texttt{[tchir]}. In all cases Fig.~\ref{fig:masses2} shows that there is no significant dependence
of the results from the details of these fits, except for a very slight fluctuation of the \texttt{[tchir][a1][420]} results for $\Ms$.
Preferring to err on the side of caution, we also discard \texttt{[tchir]} fits.\\

\vspace{3truemm}

A few general points concerning the fit analysis deserve to be highlighted:
\begin{itemize}
\item
In all our fits the $\chi^2$/dof is well below 1. This is partly because our data are correlated --- both from the fact that there are common renormalisation factors and improvement coefficients, and because we are including the contribution to the $\chi^2$ from the fluctuations of the meson masses (horizontal errors). Therefore, while the goodness-of-fit is in general satisfactory, we will refrain from quoting the corresponding p-values, since they are not really meaningful.
\item
Unsurprisingly, the inclusion of a second discretisation term $\propto \phi_2(a^2/t_0)$ in the fits contributes to an increase of the error. 
This term is often compatible with zero, and almost always so within $\sim 2\sigma$, suggesting that fits \texttt{[a1]} are safe.
As stated previously, exceptions are fits \texttt{[chi12]} and \texttt{[chipc]}, where inclusion of this term has a strong effect.
\item
Within large uncertainties, the coefficients of the leading cutoff effects  (i.e. those $\propto a^2/t_0$) depend on the fitted observable,
and are larger for $\phi_{13}$ than for $\phi_{12}$.
\item
The power-series fits \texttt{[tay12]} and \texttt{[tay13]} behave remarkably well. 
Results from \texttt{[tay12]} vanish within errors in the chiral limit,
except for fits going up to the symmetric point, which are sometimes incompatible with naught by 2–3$~\sigma$.
This is evidence that our data are not precise enough to capture the impact
of chiral logs. Fits \texttt{[tay13]} to $\phi_{lh}$ are very stable, and impressively
better than those obtained with the $\chi$PT ansatz.
Indeed, if one considers fits \texttt{[texp1]} and \texttt{[texp2]},
which are safest from the point of view of error estimation, all the
fits considered provide compatible results for $M_{13}$ within one sigma.
Notice, furthermore, that the constant terms of \texttt{[tay12]} and \texttt{[tay13]}
are generally in good agreement, signalling the consistency of the approach.
It is also interesting to note that the coefficient of the quadratic term is
very small and always compatible with zero within 1$\sigma$ (save for two cases
where it vanishes within 2$\sigma$).
\item 
Fits \texttt{[chirc]} and \texttt{[chirr]} appear to be the stablest.
\item 
NLO $\chi$PT appears to be suffering around and above $400~\MeV$.
\end{itemize}

\section{Final results and discussion}
\label{sec:final}

\begin{figure*}[t!]
\begin{center}
\hspace*{-35mm}
\begin{minipage}{.45\linewidth}
\includegraphics[width=1.5\linewidth]{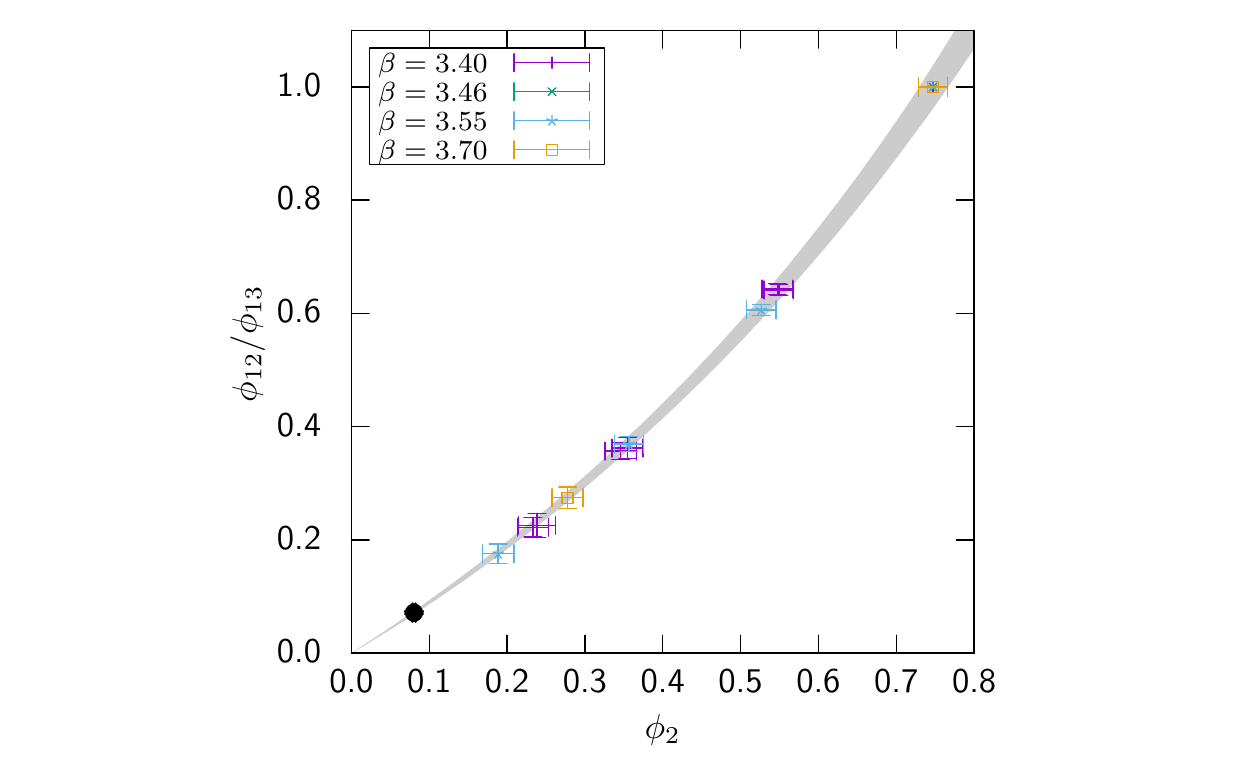}
\end{minipage}
\begin{minipage}{.45\linewidth}
\includegraphics[width=1.5\linewidth]{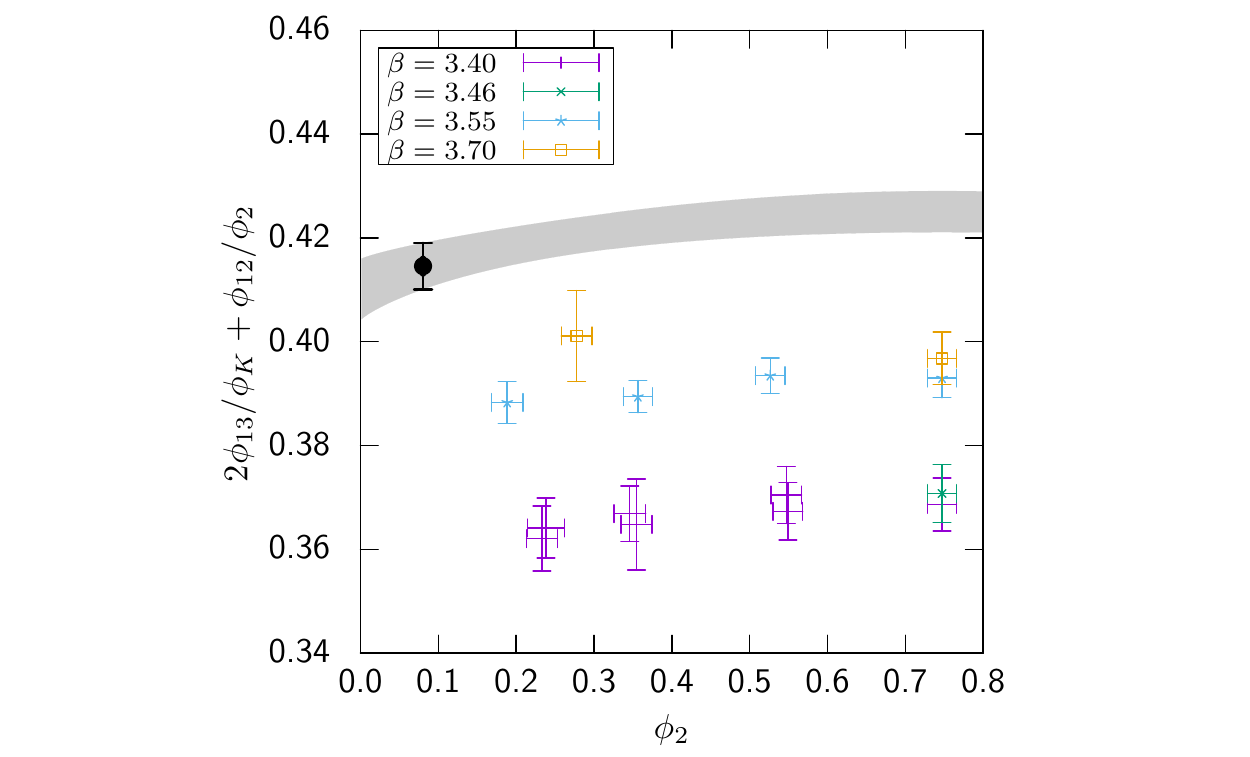}
\end{minipage}
\end{center}
\vspace{-5mm}
\caption{
Illustration of the chiral+continuum fit from which our central values are obtained.
The grey band is the continuum limit of our fit, and the full black point corresponds to
our extrapolation to the physical point.
}
\label{fig:main_fit}
\end{figure*}

Following the analysis of sect.~\ref{sec:phys-mass}, we quote as final results those obtained
from  the following procedure:
\begin{itemize}
\item The central values are those of a combined fit to the ratio $\phi_{12}/\phi_{13}$ and the quantity 
$2\phi_{13}/\phi_K+\phi_{12}/\phi_2$, using NLO $\chi$PT, with pseudoscalar meson masses
less than 360~MeV and a discretisation term proportional to $a^2/t_0$ (i.e., fit \texttt{[chirr][a1][360]}).
The error from this fit will appear as the first uncertainty in the results below.
The fit is illustrated in Fig.~\ref{fig:main_fit}.
\item We estimate systematic errors from the spread of central values of all other 
\texttt{[chirr]} and \texttt{[chirc]} fits, for all pion mass cutoffs, and for both [a1] and [a2].
The spread is intended to be the difference between the central value, obtained
as described in the previous item, and the most distant central value of all other \texttt{[chirr]}
and \texttt{[chirc]} fits. This is the second error of the results below.
Recall that \texttt{[chirc]} are combined fits to $\phi_{13}$ and the ratio $\phi_{12}/\phi_{13}$, 
using NLO $\chi$PT.
\item Discard other fits, including \texttt{[chipc]}, considered too unstable.
\item All results have been obtained using the Symanzik $\tilde b$-parameters computed in the LCP-0 case
(see discussion in sect.~\ref{sec:simul}). Using LPC-1 results instead, has very marginal effects on the error.
\item In section~\ref{sec:simul} we have also argued that for the quantities under
consideration finite volume effects are negligible.

\end{itemize}
The resulting RGI masses are
\begin{align}
&\Ms = 127.0(3.1)(3.2)~\MeV \,, \nonumber \\
&\Mud = 4.70(15)(12)~\MeV \,.
\end{align}
The quark mass ratio is obtained from
\begin{gather}
\dfrac{\Ms}{\Mud} = \dfrac{2}{\phi_{ll}/\phi_{lh}}-1\,.
\end{gather}
Dependence on renormalisation is only implicit, from the joint fit with $\phi_{13}$.
The same procedures as above yield
\begin{gather}
\dfrac{\Ms}{\Mud} = 27.0(1.0)(0.4) \,.
\label{eq:MRGI_rato}
\end{gather}
The above results for RGI masses refer to the $\nf=2+1$ theory.

It is customary in phenomenological studies to report light quark masses measured
in the $\nf=2+1$ lattice theory in the $\msbar$ scheme at 2~GeV, referred to the more physical 
QCD with four flavours. This entails using $\nf=3$ perturbative RG-running from 2~GeV
down to the charm threshold, followed by $\nf=4$ perturbative RG-running back to 2~GeV;
see for example ref.~\cite{Aoki:2019cca}. We use 4-loop perturbative RG-running
and the value for the $\Lambda_{\rm QCD}^{\msbar}$ parameter computed by the
ALPHA Collaboration in ref.~\cite{Bruno:2017gxd} to obtain\footnote{In converting
our results to $\msbar$ we have taken into account the uncertainty in the matching
factor coming from the error on $\Lambda_{\rm QCD}^{\msbar}$,
as well as the covariance of the latter with our determination of $M_i$.}
\begin{align}
&m_{\mathrm s \rm R}(2~{\rm GeV})   = 95.7(2.5)(2.4)~\MeV \,, \nonumber \\
&m_{\mathrm{u/d} \rm R}(2~{\rm GeV}) = 3.54(12)(9)~\MeV \,.
\end{align}
The mass ratio is obviously the same as in eq.~(\ref{eq:MRGI_rato}).
We note in passing that switching to the four-flavour theory has a very small effect on $\msbar$ results,
since at $2~{\rm GeV}$ the matching factor is $m_{\rm R}(\nf=4)/m_{\rm R}(\nf=3) =1.002$.

\begin{figure*}[t!]
\begin{center}
\includegraphics[width=0.8\textwidth]{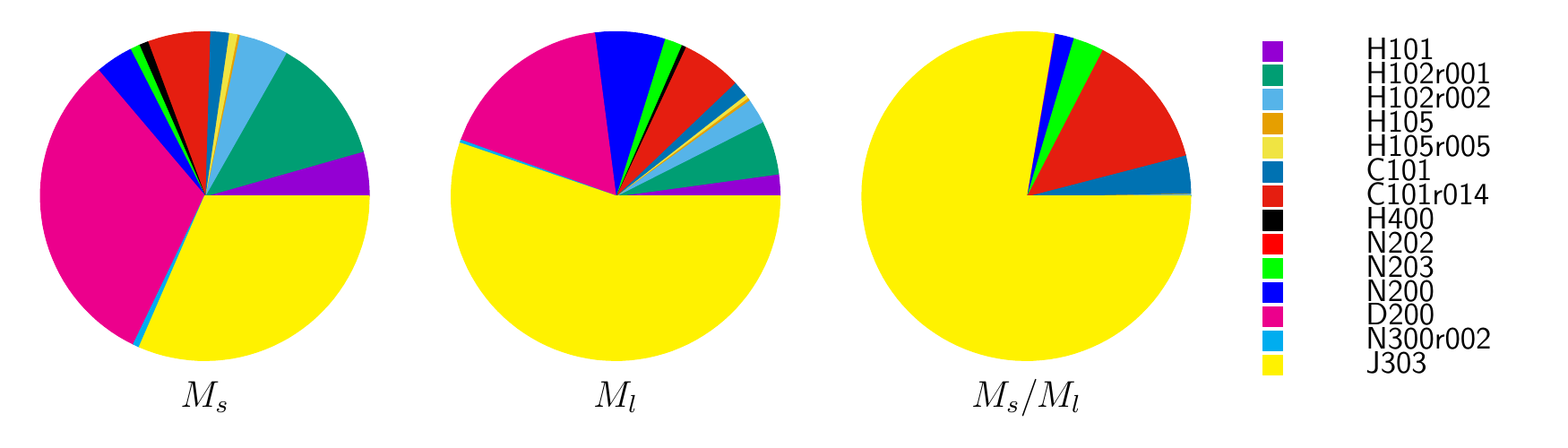}
\end{center}
\vspace{-5mm}
\caption{
Contributions to the statistical+chiral extrapolation+continuum limit uncertainties from each ensemble included in our analysis,
for our preferred fit \texttt{[chirr][a1][360]}.
}
\label{fig:error_budget}
\end{figure*}

The error budget for our computation is summarised in Table~\ref{tab:error_budget} and Fig~\ref{fig:error_budget}.
Uncertainties are completely dominated by our chiral fits. We have separated these errors into two contributions;
see first two lines of Table~\ref{tab:error_budget}. The first error is that of our best fit \texttt{[chirr][a1][360]},
and includes the statistical errors as well as the error from combined fits in $\phi_2$ and $a$. The second
uncertainty is the one arising upon varying the fit ans\"atze and their $\phi_2$ range. All other errors are
clearly seen to be subdominant.
It is worth noting that, as expected, the largest contribution to the uncertainty comes from the ensembles
with the lightest sea pion masses, especially the one with the finest lattice spacing. It is then clear that
decreasing our errors would require more chiral ensembles, and more extensive simulations at light masses.

\begin{table*}
\caption{
Contributions to the squared errors of our final quantities from different sources.
}
\label{tab:error_budget}
\begin{tabular*}{\textwidth}{@{\extracolsep{\fill}}lrrr@{}}
\hline\noalign{\smallskip}
 & \multicolumn{1}{c}{$\Mud$} & \multicolumn{1}{c}{$\Ms$} & \multicolumn{1}{c}{$\Ms/\Mud$} \\
\noalign{\smallskip}\hline\noalign{\smallskip}
stat+chiral+cont  &     $56\%$   &     $40\%$   &   $86\%$     \\
fit systematics   &     $39\%$   &     $52\%$   &   $14\%$     \\
renormalisation   &     $<1\%$   &     $<1\%$   &     n/a      \\
running           &      $5\%$   &      $8\%$   &     n/a      \\
${\rm O}(a)$ impr & (negligible) & (negligible) & (negligible) \\
finite volume     & (negligible) & (negligible) & (negligible) \\
\noalign{\smallskip}\hline
\end{tabular*}
\end{table*}

The current FLAG~2019~\cite{Aoki:2019cca} world averages from $\nf =2+1$ simulations, in the $\msbar$ scheme, reportedly quoted for 
the $\nf=4$ theory as explained above, are:
\begin{align}
&m_{\mathrm s \rm R}(2 {\rm GeV}) = 92.03(88) \MeV \,, \nonumber \\
&m_{\mathrm{u/d} \rm R}(2 {\rm GeV}) = 3.364(41) \MeV \,.
\end{align}
The strange mass estimate is based on the results of refs.~\cite{Blum:2014tka,Durr:2010vn,Durr:2010aw,McNeile:2010ji,Maezawa:2016vgv,Bazavov:2009fk},
while the up/down one is based on refs.~\cite{Blum:2014tka,Durr:2010vn,Durr:2010aw,McNeile:2010ji,Bazavov:2010yq}. For the quark mass ratio, based on
refs.~\cite{Blum:2014tka,Durr:2010vn,Durr:2010aw,Bazavov:2009fk}, FLAG quotes
\begin{equation}
\dfrac{m_{\mathrm s \rm R}}{m_{\mathrm{u/d} \rm R}} = 27.42(12) \MeV \,\, .
\end{equation}
Our results for the strange and light quark masses agree with those of FLAG within $1.7\sigma$
and $1.2\sigma$ respectively and thus exhibit good compatibility albeit with bigger errors.

\begin{acknowledgement}%
We thank T.~Blum, G.~Herdo\'{\i}za, L.~Lellouch and A.~Portelli for useful discussions.
J.K., C.P., and A.V. thank CERN for its hospitality. A.V. also thanks BNL for its hospitality.
C.P. and D.P. thankfully acknowledge support through the Spanish projects
FPA2015-68541-P (MINECO/FEDER) and PGC2018-094857-B-I00, the Centro de Excelencia Severo Ochoa
Programme SEV-2016-0597, and the EU H2020-MSCAITN-2018-813942 (EuroPLEx).
We acknowledge PRACE for awarding us access to resource FERMI based in Italy
at CINECA, Bologna, and to resource SuperMUC based in Germany at LRZ, Munich.
Computer resources were also provided by the INFN (MARCONI cluster at CINECA, Bologna);
a grant from the Swiss National Supercomputing Centre (CSCS) under project ID s38;
the Gauss Centre for Supercomputing (GCS) --- through the John von Neumann Institute for Computing
(NIC) --- on the GCS share of the supercomputer JUQUEEN at Jülich Supercomputing Centre (JSC);
Altamira, provided by IFCA at the University of Cantabria;
the FinisTerraeII machine provided by CESGA (Galicia Supercomputing Centre);
and a dedicated HPC cluster at CERN.
We are grateful for the technical support received by all the computer centers.
\end{acknowledgement}

\appendix
\section{Chiral perturbation theory expansions}
\label{app-chipt}

We adapt standard $\chi$PT expressions to our specific parametrisation of the data, stemming from our choice of chiral trajectory. We start, for example, from eqs.~(B5) and (B6) of ref.~\cite{Allton:2008pn}, which are NLO chiral expansions of the light and strange pseudoscalar mesons $m_{\pi}$ and $m_{\rm K}$ in terms of light and strange quark masses $m_1 = m_2$ and $m_3$. These series are inverted, so that quark masses are functions of meson masses. The PCAC quark mass combinations $m_{12} = (m_1 + m_2)/2$ and $m_{13} = (m_1 + m_3)/2$ are then formed and everything is  re-expressed in terms of the dimensionless quark masses $\phi_{\rm 12}$, $\phi_{\rm 13}$ and the  dimensionless quantities $\phi_2$ and $\phi_4$, so that we arrive at
\begin{eqnarray}
  \phi_{12} &=& \dfrac{\phi_2}{2 B_0 \sqrt{8t_0}} \,\cdot\,\Bigg\{\, 1  
  \, - \, \dfrac{16}{8t_0 f_0^2}(2 L_8 - L_5) \phi_2 \nonumber \\
  &-&\, \dfrac{32}{8t_0 f_0^2}(2L_6-L_4) \phi_4 \, -\, \dfrac{1}{24\pi^2 8 t_0 f_0^2}\times \\
  &\,& \bigg[\dfrac{3}{2} \phi_2\ln\bigg ( \dfrac{\phi_2}{8 t_0 \Lambda_\chi^2} \bigg ) 
\, - \,\dfrac{1}{2}\phi_\eta\ln \bigg ( \dfrac{\phi_\eta}{8 t_0 \Lambda_\chi^2}\bigg ) \bigg] \,\Bigg\} \,\, ,\nonumber
\label{eq:chPTsu3:ml_phi}
\end{eqnarray}
and
\begin{eqnarray}
\phi_{13} &=& \dfrac{2\phi_4 - \phi_2}{4 B_0 \sqrt{8t_0}} \,\cdot\,\Bigg\{\,1\,
-\,\dfrac{8}{8t_0 f_0^2}(- 2 L_8 + L_5) \phi_2\nonumber \\
&-&\,\dfrac{16}{8t_0 f_0^2}(4L_6 + 2 L_8 -2L_4 - L_5) \phi_4 \\
&-& \dfrac{1}{24\pi^2 8t_0 f_0^2}\,\phi_\eta\ln \bigg ( \dfrac{\phi_\eta}{8t_0 \Lambda_\chi^2}\bigg ) \,\Bigg\} \,\, ,\nonumber
\label{eq:chPTsu3:mh+ml_phi}
\end{eqnarray}
where $B_0, f_0, L_k (k=4,5,6,8)$ are standard $\chi$PT parameters and
\begin{equation}
\label{eq:def=phieta}
\phi_\eta \,\, \equiv 8 t_0 \dfrac{4 m_{\rm K}^2 - m_\pi^2}{3} \,\, = \,\, \dfrac{4 \phi_4 - 3 \phi_2}{3}  \,\, .
\end{equation}
The NLO LECs $L_i$ and $B_0$ are implicitly renormalised at scale $\Lambda_\chi$.  It is also useful to consider the ratio
\begin{align}
\label{eq:philldphilhchipt}
& \dfrac{\phi_{12}}{\phi_{13}} = \dfrac{2\phi_2}{2\phi_4 - \phi_2} \Bigg\{
1 - \dfrac{24}{8t_0f_0^2}(2L_8-L_5)\left[\phi_2 - \dfrac{2}{3}\phi_4\right] \nonumber \\
&  -
\dfrac{1}{16\pi^2(8t_0f_0^2)}\left[
\phi_2\ln\left(\dfrac{\phi_2}{8t_0\Lambda_\chi^2}\right) -
\phi_\eta\ln\left(\dfrac{\phi_\eta}{8t_0\Lambda_\chi^2}\right)
\right]
\Bigg\}\,\, ,
\end{align}
which does not require renormalisation. Note that at the symmetric point ($\phi_2  = 2 \phi_4 /3$) the current quark masses of eqs.~(\ref{eq:chPTsu3:ml_phi}) and (\ref{eq:chPTsu3:mh+ml_phi}) respect the constraint $\phi_{12} = \phi_{13}$, while the ratio~(\ref{eq:philldphilhchipt}) is exactly~1. Note that the sum of ratios 
\begin{align}
\label{eq:pregolden}
&\dfrac{4 \phi_{13}}{2\phi_4-\phi_2} + \dfrac{\phi_{12}}{\phi_2}  = \dfrac{3}{2B_0\sqrt{8t_0}}\times \\
&\Bigg\{1 - \dfrac{16}{8t_0f_0^2}\left(\dfrac{4}{3}L_8-\dfrac{2}{3}L_5+4L_6-2L_4\right)\phi_4 \nonumber \\
& - \dfrac{1}{48\pi^2(8t_0f_0^2)}\left[
\phi_2\ln\left(\dfrac{\phi_2}{8t_0\Lambda_\chi^2}\right) +
\phi_\eta\ln\left(\dfrac{\phi_\eta}{8t_0\Lambda_\chi^2}\right) \right]
\Bigg\}\,\, ,\nonumber
\end{align}
has the remarkable advantages of depending on just one combination of NLO LECs, and of being free of polynomial dependence on $\phi_2$. 

We next rewrite eqs.~(\ref{eq:chPTsu3:ml_phi}) and (\ref{eq:chPTsu3:mh+ml_phi}) in forms which are suitable for combined fits, with common  coefficients for $\phi_{12}$ and $\phi_{13}$, obtaining
\begin{align}
\label{eq:phill_chipt}
\phi_{12} & = \phi_2\left[p_1 + p_2\phi_2 + p_3 K\left(\mathcal{L}_2 - \dfrac{1}{3}\mathcal{L}_\eta\right)\right]
\,\, ,\\
\label{eq:philh_chipt}
\phi_{13} & = \dfrac{2\phi_4 - \phi_2}{2}\left[p_1 + p_2 \left(\phi_4 - \dfrac{\phi_2}{2} \right) + \dfrac{2}{3}p_3 K\mathcal{L}_\eta\right]
 \,\, ,
\end{align}
where the coefficients $p_1, p_2$, and $p_3$ relate to LECs as follows:
\begin{align}
\label{eq:p1}
p_1 & = \dfrac{1}{2B_0\sqrt{8t_0}}\left[1-\dfrac{32}{8t_0f_0^2}(2L_6-L_4)\phi_4\right] \nonumber \\
&\approx \dfrac{1}{2B_0\sqrt{8t_0}}\left[1-\dfrac{32}{8t_0f_{\pi K}^2}(2L_6-L_4)\phi_4\right]\,\, , \\
\label{eq:p2}
p_2 & = -\dfrac{1}{2B_0\sqrt{8t_0}}\dfrac{16}{8t_0f_0^2}(2L_8-L_5)\nonumber \\
&\approx  -\dfrac{1}{2B_0\sqrt{8t_0}}\dfrac{16}{8t_0f_{\pi K}^2}(2L_8-L_5)\,\, , \\
\label{eq:p3}
p_3 & = -\dfrac{1}{2B_0\sqrt{8t_0}} \,\, \,.
\end{align}
We also define
\begin{equation}
\label{eq:K}
K \equiv (8t_016\pi^2 f_0^2)^{-1} \approx (8t_016\pi^2f_{\pi K}^2)^{-1} \,\, ,
\end{equation} 
with $f_{\rm \pi K}$ given by eq.~(\ref{eq:fpiK}). The chiral logarithms are
\begin{align}
\label{chilog-2}
\mathcal{L}_2 \equiv  \phi_2\ln\phi_2 \,\, , \quad
%\label{chilog-eta}
\mathcal{L}_\eta \equiv \phi_\eta\ln\phi_\eta \,\, . 
\end{align}
The following points should be kept in mind:
\begin{itemize}
\item We are using only configurations along the $\phi_4={\rm constant}$ chiral trajectory. Terms proportional to $\phi_4$ are thus reabsorbed into constant fit terms.
\item Our expressions are linear in fit parameters, rather than non-linear factors in which LECs appear explicitly. Determination of LECs is beyond the scope of the present work.
\item By replacing $f_0^2$ by $f_{\pi K}^2$ in the above definition of $K$, the coefficients of chiral logarithms are completely fixed relative to the LO value; cf. eqs.~(\ref{eq:chPTsu3:ml_phi}) and (\ref{eq:chPTsu3:mh+ml_phi}). This eliminates one fit parameter, pushing its effect to NNLO LECs. In practice, the fact that terms with $\phi_4$ are reabsorbed into the LO terms nullifies the effect in some fits, e.g., those for $\phi_{12}$ and $\phi_{13}$. A second advantage of this choice is that the resulting ans\"atze are fully linear in the fit parameters. See also eq. (2.5) in \cite{Bruno:2016plf} and comments therein on the reasons that $f_0 \approx f_{\pi K}$ and  for preferring $f_{\pi K}$ to $f_0$.
\item We conveniently set the renormalisation scale to $\Lambda_\chi=1/\sqrt{8t_0} \simeq 476~\MeV$, simplifying  the chiral logs. There is no need to reabsorb $\ln(8t_0\Lambda_\chi^2)$ terms in fit parameters. This is an unconventional choice, as common practice consists in providing results for LECs at $\Lambda_\chi=m_\rho$ or $\Lambda_\chi=4\pi f_0$. Consequently, NLO LECs eventually obtained with our methodology may only be compared to results in the literature after some extra work.
\end{itemize}

Using the above expressions and consistently neglecting higher mass orders, we obtain for the ratio~(\ref{eq:philldphilhchipt}) of PCAC masses
\begin{gather}
\label{eq:phirat_chipt}
\dfrac{\phi_{12}}{\phi_{13}} = \dfrac{2\phi_2}{2\phi_4 - \phi_2}\left[1 + \dfrac{p_2}{p_1} \left(\dfrac{3}{2}\phi_2 - \phi_4\right) - \tilde K\left(\mathcal{L}_2-\mathcal{L}_\eta\right)\right]
\,\, .
\end{gather}
For the combination~(\ref{eq:pregolden}) we have
\begin{gather}
\label{eq:golden_chipt}
\dfrac{4\phi_{13}}{2\phi_4-\phi_2} + \dfrac{\phi_{12}}{\phi_2} = 3p_1 + 2p_2\phi_4 + p_3 K\left(\mathcal{L}_2+\mathcal{L}_\eta\right) \,\, .
\end{gather}

With $\phi_4$ held constant, the quantities of eqs.~(\ref{eq:phill_chipt}), (\ref{eq:philh_chipt}), (\ref{eq:phirat_chipt}), and (\ref{eq:golden_chipt}) are functions of $\phi_2$ only. We use these expressions to fit our data, after adding $\rmO(a^2)$ terms which model leading discretisation effects that have been neglected throughout this Appendix.

\section{Discretisation effects}
\label{sec:discretisation}

In order to parametrise the discretisation effects of the quantities we fit, we first examine $\phi_{ij}$; cf. eqs.~(\ref{eq:renmassPCAC}) and (\ref{eq:defphiij}). It can be written in the very general form
\begin{equation}
\phi_{ij}=\phi_{ij}^{\textrm{cont}}+f\big(a,\frac{m_i+m_j}{2},\frac{m_i-m_j}{2}, \tr [ \Mq] \big),
\end{equation}
where $\phi_{ij}^{\textrm{cont}}$ is the continuum quantity and the function $f$ contains the discretisation effects which in general depend on the lattice spacing $a$, the quark masses $m_i$, $m_j$, and the trace of the mass matrix $\tr [\Mq]$. As we have discussed in Section~\ref{sec:genth}, we will ignore $\rmO(g_0^4 \Tr [M_{\rm sum}])$ discretisation effects and only
consider the influence of $\rmO(a^2)$ uncertainties.
Also $\phi_{ij}$ has to be symmetric with respect to the exchange of quarks, $i \leftrightarrow j$. We can thus parametrise $f$ as follows:
\begin{align}
  &f\big(a,\frac{m_i+m_j}{2},\frac{m_i-m_j}{2}, \tr [ \Mq ]\big) =\nonumber \\
  &c_0\dfrac{a^2}{t_0}+c_1\dfrac{a^2}{t_0}\sqrt{8t_0}\bigg(\dfrac{m_i+m_j}{2}\bigg) + c_2 \dfrac{a^2}{t_0} \sqrt{8t_0} \tr [M_{\rm _q}] \nonumber \\
  & +c_3\dfrac{a^2}{t_0}8t_0\bigg(\dfrac{m_i+m_j}{2}\bigg)^2 +c_4\dfrac{a^2}{t_0}8t_0\bigg(\dfrac{m_i-m_j}{2}\bigg)^2 \nonumber \\
  &+ c_5 \dfrac{a^2}{t_0} 8t_0 \tr[\Mq^2] + c_6 \dfrac{a^2}{t_0} 8t_0 (\tr [\Mq])^2 \nonumber \\
  &+ c_7 \dfrac{a^2}{t_0}\sqrt{8t_0}\bigg(\dfrac{m_i+m_j}{2} \bigg) \sqrt{8t_0} \tr [\Mq] + \rmO(a^3).
  \label{eq:f-general}
\end{align}
A further simplification is brought about by neglecting the dependence of $c_0, \ldots , c_7$ on the bare coupling $g_0^2$.

Next we write the function $f$ in terms of $\phi_2$, recalling that a constant $\phi_4$ constrains the relation between the heavier (strange) and light quark masses.  This is done by first expressing the current quark masses on the rhs of the above equation in terms of $\phi_{12}$ and $\phi_{13}$, followed by using their LO \chiPT~relations to $\phi_2$ and $\phi_4$. In particular, with $\beta_0 \equiv 1/(2 B_0 \sqrt{8 t_0})$, we see from eqs.~(\ref{eq:phill_chipt}), (\ref{eq:philh_chipt}) that to LO:
\begin{align}
\label{eq:phi12-LO}
&\phi_{12} \stackrel{\mathclap{\normalfont\tiny\mbox{LO}}}{=}  \beta_0\phi_2 \,\, , \\
\label{eq:phi13-LO}
&\phi_{13} \stackrel{\mathclap{\normalfont\tiny\mbox{LO}}}{=}  \beta_0\dfrac{1}{2}(2\phi_4-\phi_2) \,\, , \\
\label{eq:m1-m3-LO}
&\sqrt{8t_0} \bigg(\dfrac{m_1-m_3}{2}\bigg) \phi_{12}-\phi_{13} \stackrel{\mathclap{\normalfont\tiny\mbox{LO}}}{=}  \beta_0\bigg(\dfrac{3}{2}\phi_2-\phi_4\bigg) \,\, , \\
\label{eq:tr-LO}
&\sqrt{8t_0} \tr [\Mq]  =   \sqrt{8t_0} [ 2 m_1 + m_3] \,\, \stackrel{\mathclap{\normalfont\tiny\mbox{LO}}}{=} \,\, 2 \beta_0 \phi_4 \,\, , \\
&8t_0 \tr [\Mq^2]  =   8t_0 [ 2 m_1^2 + m_3^2] \,\, = \,\,  2 \phi_{12}^2 + \phi_h^2  \nonumber \\
&\qquad\qquad\,\,\, \stackrel{\mathclap{\normalfont\tiny\mbox{LO}}}{=}  \,\,  2 \beta_0^2 [ 3 \phi_2^2 + 2\phi_4^2 - 4 \phi_4 \phi_2] \,\, .
\label{eq:trsq-LO}
\end{align}

Inserting the above LO expressions in eq.~(\ref{eq:f-general}) we obtain, after some straightforward algebra, that for two light quarks the discretisation function has the form
\begin{align}
\label{eq:fll}
f_{12}(a,\phi_2) & \equiv f(a, m_1, 0,\tr [\Mq]) \nonumber \\
&= \dfrac{a^2}{8 t_0} \big [ C_0+C_1 \phi_2+C_2 \phi_2^2 \big ]+\rmO(a^3) \,\, ,
\end{align}
where $C_k$ (with $k = 0, 1, \dots $) depend on the constants $\beta_0$, $\phi_4$, and the coefficients $c_l$, suitably rescaled by factors of $8 t_0$ (with $l = 0, 1, \dots $). Similarly,
for the heavier and a light quark we obtain
\begin{align}
\label{eq:flh}
f_{13}(a,\phi_2)  &\equiv  f(a, m_1, m_3, \tr [\Mq]) \nonumber \\
&= \dfrac{a^2}{8 t_0} \big [ \widetilde{C}_0 +\widetilde{C}_1\phi_2+\widetilde{C}_2\phi_2^2\big ] +\rmO(a^3). 
\end{align}
Note that, although in general coefficients $C_n$ and $\widetilde{C}_n$ are not the same, in the case of $m_3  = m_1$ (symmetric point) $f_{13} = f_{12}$ trivially.

The very fact that we have used LO $\chi$PT to obtain the last two expressions (cf. eqs.~(\ref{eq:phi12-LO})-(\ref{eq:trsq-LO})) allows us to drop $\rmO(a^2 \phi_2^2)$ contributions of $f_{12}$ and $f_{13}$. Moreover, standard power-counting schemes in Wilson $\chi$PT~\cite{Rupak:2002sm,Bar:2003mh} suggest that terms of $\rmO(a^2)$ enter at the same order as $\rmO(m_\pi^2)$, which would imply that the terms of $\rmO(a^2 \phi_2)$ should also be dropped. We will nevertheless keep this term and explore its impact.

For the ratio of $\phi_{12}$ and $\phi_{13}$ we have that
\begin{align}
  \dfrac{\phi_{12}}{\phi_{13}} &= \dfrac{\phi_{12}^{\textrm{cont}}+f_{12}(a,\phi_2)}{\phi_{13}^{\textrm{cont}}+f_{13}(a,\phi_2)}
  \nonumber \\
&= \dfrac{\phi_{12}^{\textrm{cont}}}{\phi_{13}^{\textrm{cont}}} + \dfrac{f_{12}}{\phi_{13}^{\textrm{cont}}} - \dfrac{f_{13} \phi_{12}^{\textrm{cont}}}{(\phi_{13}^{\textrm{cont}})^2} +\cdots \,\,\, .
\end{align}
We write the discretisation functions $f_{12}$ and $f_{13}$ as in eqs.~(\ref{eq:fll}) and~(\ref{eq:flh}) and then express coefficients $C_0, C_1, \widetilde{C}_0, \widetilde{C}_1, \dots$ in terms of the original coefficients $c_i$ of eq.~(\ref{eq:f-general}). After some algebra we end up with
\begin{align}
\dfrac{\phi_{12}}{\phi_{13}} &=  \dfrac{\phi_{12}^{\textrm{cont}}}{\phi_{13}^{\textrm{cont}}} + \dfrac{a^2}{8t_0} \dfrac{2\phi_4 - 3\phi_2}{(2 \phi_4- \phi_2)^2} \Big [ D_0 + D_1\phi_2 + D_2 \phi_2^2 \Big ]\nonumber \\ &+ \rmO(a^3) \,\, .
\label{eq:ratio-12-13-discr}
\end{align}
The coefficients $D_0, D_1, D_2, \dots$ depend on the $c_i$'s. The factor $2 \phi_4 - \phi_2$ in the discretisation term vanishes at the symmetric point $\phi_2= 2 \phi_4/3$. This confirms that at the symmetric point the ratio $\phi_{12}/\phi_{13}$ is 1 by construction, for any lattice spacing. In analogy to the arguments exposed above for $f_{12}$ and $f_{13}$, we drop the $D_2 \phi_2^2$ term in our fits. Moreover, the variation of the denominator $(2\phi_4 - \phi_2)^2$ is relatively mild, ranging between $\sim 2$ and $\sim 4.6$ as  $\phi_2$ varies between $\sim 0.1$ and $\sim 0.8$ in our simulations. To simplify matters, we reabsorb this $\rmO(1)$ term in re-definitions of $D_0$ and $D_1$.

Finally, for the combination of eq.~(\ref{eq:golden_chipt}), we straightforwardly parametrise the discretisation errors in a way analogous to $f_{12}$ and $f_{13}$; see eq.~(\ref{eq:golden_chipt_disc}).

\small
\addcontentsline{toc}{section}{References}
\bibliographystyle{JHEP}
\bibliography{lattice}

\providecommand{\href}[2]{#2}\begingroup\raggedright\begin{thebibliography}{10}

\bibitem{Symanzik:1973vg}
K.~Symanzik, \emph{{Infrared singularities and small distance behavior
  analysis}}, \href{https://doi.org/10.1007/BF01646540}{\emph{Commun. Math.
  Phys.} {\bfseries 34} (1973) 7}.

\bibitem{Appelquist:1974tg}
T.~Appelquist and J.~Carazzone, \emph{{Infrared Singularities and Massive
  Fields}}, \href{https://doi.org/10.1103/PhysRevD.11.2856}{\emph{Phys. Rev.}
  {\bfseries D11} (1975) 2856}.

\bibitem{Ovrut:1980uv}
B.~A. Ovrut and H.~J. Schnitzer, \emph{{Gauge Theories With Minimal Subtraction
  and the Decoupling Theorem}},
  \href{https://doi.org/10.1016/0550-3213(81)90011-0}{\emph{Nucl. Phys.}
  {\bfseries B179} (1981) 381}.

\bibitem{Bernreuther:1981sg}
W.~Bernreuther and W.~Wetzel, \emph{{Decoupling of Heavy Quarks in the Minimal
  Subtraction Scheme}}, \href{https://doi.org/10.1016/0550-3213(82)90288-7,
  10.1016/S0550-3213(97)00811-0}{\emph{Nucl. Phys.} {\bfseries B197} (1982)
  228}.

\bibitem{Aoki:2019cca}
{S. Aoki et al. (Flavour Lattice Averaging Group)}, \emph{{FLAG Review 2019}},
  \href{https://arxiv.org/abs/1902.08191}{{\ttfamily 1902.08191}}.

\bibitem{Knechtli:2017xgy}
{\scshape ALPHA} collaboration, F.~Knechtli, T.~Korzec, B.~Leder and G.~Moir,
  \emph{{Power corrections from decoupling of the charm quark}},
  \href{https://doi.org/10.1016/j.physletb.2017.10.025}{\emph{Phys. Lett.}
  {\bfseries B774} (2017) 649}
  [\href{https://arxiv.org/abs/1706.04982}{{\ttfamily 1706.04982}}].

\bibitem{Bruno:2014jqa}
M.~Bruno et~al., \emph{{Simulation of QCD with N$_{f} =$ 2 $+$ 1 flavors of
  non-perturbatively improved Wilson fermions}},
  \href{https://doi.org/10.1007/JHEP02(2015)043}{\emph{JHEP} {\bfseries 02}
  (2015) 043} [\href{https://arxiv.org/abs/1411.3982}{{\ttfamily 1411.3982}}].

\bibitem{Bruno:2016plf}
M.~Bruno, T.~Korzec and S.~Schaefer, \emph{{Setting the scale for the CLS $2 +
  1$ flavor ensembles}},
  \href{https://doi.org/10.1103/PhysRevD.95.074504}{\emph{Phys. Rev.}
  {\bfseries D95} (2017) 074504}
  [\href{https://arxiv.org/abs/1608.08900}{{\ttfamily 1608.08900}}].

\bibitem{Wilson}
K.~G. Wilson, \emph{Confinement of quarks}, {\emph{Phys. Rev.} {\bfseries D10}
  (1974) 2445}.

\bibitem{Sheikholeslami:1985ij}
B.~Sheikholeslami and R.~Wohlert, \emph{{Improved Continuum Limit Lattice
  Action for QCD with Wilson Fermions}},
  \href{https://doi.org/10.1016/0550-3213(85)90002-1}{\emph{Nucl. Phys.}
  {\bfseries B259} (1985) 572}.

\bibitem{Luscher:1984xn}
M.~{L\"uscher} and P.~Weisz, \emph{{On-Shell Improved Lattice Gauge Theories}},
  \href{https://doi.org/10.1007/BF01206178}{\emph{Commun. Math. Phys.}
  {\bfseries 97} (1985) 59}.

\bibitem{Bulava:2013cta}
J.~Bulava and S.~Schaefer, \emph{{Improvement of $N_f=3$ lattice QCD with
  Wilson fermions and tree-level improved gauge action}},
  \href{https://doi.org/10.1016/j.nuclphysb.2013.05.019}{\emph{Nucl. Phys.}
  {\bfseries B874} (2013) 188}
  [\href{https://arxiv.org/abs/1304.7093}{{\ttfamily 1304.7093}}].

\bibitem{Campos:2018ahf}
{\scshape ALPHA} collaboration, I.~Campos, P.~Fritzsch, C.~Pena, D.~Preti,
  A.~Ramos and A.~Vladikas, \emph{{Non-perturbative quark mass renormalisation
  and running in $N_f=3$ QCD}},
  \href{https://doi.org/10.1140/epjc/s10052-018-5870-5}{\emph{Eur. Phys. J.}
  {\bfseries C78} (2018) 387}
  [\href{https://arxiv.org/abs/1802.05243}{{\ttfamily 1802.05243}}].

\bibitem{Bulava:2015bxa}
{\scshape ALPHA} collaboration, J.~Bulava, M.~Della~Morte, J.~Heitger and
  C.~Wittemeier, \emph{{Non-perturbative improvement of the axial current in
  $N_f =3$ lattice QCD with Wilson fermions and tree-level improved gauge
  action}}, \href{https://doi.org/10.1016/j.nuclphysb.2015.05.003}{\emph{Nucl.
  Phys.} {\bfseries B896} (2015) 555}
  [\href{https://arxiv.org/abs/1502.04999}{{\ttfamily 1502.04999}}].

\bibitem{Bulava:2016ktf}
J.~Bulava, M.~Della~Morte, J.~Heitger and C.~Wittemeier, \emph{{Nonperturbative
  renormalization of the axial current in $N_f = 3$ lattice QCD with Wilson
  fermions and a tree-level improved gauge action}},
  \href{https://doi.org/10.1103/PhysRevD.93.114513}{\emph{Phys. Rev.}
  {\bfseries D93} (2016) 114513}
  [\href{https://arxiv.org/abs/1604.05827}{{\ttfamily 1604.05827}}].

\bibitem{deDivitiis:2017vvw}
{\scshape ALPHA} collaboration, G.~M. de~Divitiis, M.~Firrotta, J.~Heitger,
  C.~C. {K\"oster} and A.~Vladikas, \emph{{Non-perturbative determination of
  improvement $b$-coefficients in $N_f=3$}},
  \href{https://doi.org/10.1051/epjconf/201817510008}{\emph{EPJ Web Conf.}
  {\bfseries 175} (2018) 10008}
  [\href{https://arxiv.org/abs/1710.07020}{{\ttfamily 1710.07020}}].

\bibitem{Aoki:2016frl}
S.~Aoki et~al., \emph{{Review of lattice results concerning low-energy particle
  physics}}, \href{https://doi.org/10.1140/epjc/s10052-016-4509-7}{\emph{Eur.
  Phys. J.} {\bfseries C77} (2017) 112}
  [\href{https://arxiv.org/abs/1607.00299}{{\ttfamily 1607.00299}}].

\bibitem{Luscher:2010iy}
M.~{L\"uscher}, \emph{{Properties and uses of the Wilson flow in lattice QCD}},
  \href{https://doi.org/10.1007/JHEP08(2010)071,
  10.1007/JHEP03(2014)092}{\emph{JHEP} {\bfseries 08} (2010) 071}
  [\href{https://arxiv.org/abs/1006.4518}{{\ttfamily 1006.4518}}].

\bibitem{Sommer:2014mea}
R.~Sommer, \emph{{Scale setting in lattice QCD}},
  \href{https://doi.org/10.22323/1.187.0015}{\emph{PoS} {\bfseries LATTICE2013}
  (2014) 015} [\href{https://arxiv.org/abs/1401.3270}{{\ttfamily 1401.3270}}].

\bibitem{Bar:2013ora}
O.~{B\"ar} and M.~Golterman, \emph{{Chiral perturbation theory for gradient
  flow observables}}, \href{https://doi.org/10.1103/PhysRevD.89.099905,
  10.1103/PhysRevD.89.034505}{\emph{Phys. Rev.} {\bfseries D89} (2014) 034505}
  [\href{https://arxiv.org/abs/1312.4999}{{\ttfamily 1312.4999}}].

\bibitem{Bruno:2019xed}
{\scshape ALPHA} collaboration, M.~Bruno, I.~Campos, J.~Koponen, C.~Pena,
  D.~Preti, A.~Ramos et~al., \emph{{Light and strange quark masses from
  $N_f=2+1$ simulations with Wilson fermions}},
  \href{https://doi.org/10.22323/1.334.0220}{\emph{PoS} {\bfseries LATTICE2018}
  (2019) 220} [\href{https://arxiv.org/abs/1903.04094}{{\ttfamily
  1903.04094}}].

\bibitem{Bhattacharya:2005rb}
T.~Bhattacharya, R.~Gupta, W.~Lee, S.~R. Sharpe and J.~M. Wu, \emph{{Improved
  bilinears in lattice QCD with non-degenerate quarks}},
  \href{https://doi.org/10.1103/PhysRevD.73.034504}{\emph{Phys.Rev.} {\bfseries
  D73} (2006) 034504} [\href{https://arxiv.org/abs/hep-lat/0511014}{{\ttfamily
  hep-lat/0511014}}].

\bibitem{deDivitiis:2019xla}
{\scshape ALPHA} collaboration, G.~M. Divitiis, P.~Fritzsch, J.~Heitger, C.~C.
  K{\"o}ster, S.~Kuberski and A.~Vladikas, \emph{{Non-perturbative
  determination of improvement coefficients $b_\mathrm{m}$ and
  $b_\mathrm{A}-b_\mathrm{P}$ and normalisation factor
  $Z_\mathrm{m}Z_\mathrm{P}/Z_\mathrm{A}$ with $N_\mathrm{f}= 3$ Wilson
  fermions}}, \href{https://doi.org/10.1140/epjc/s10052-019-7287-1}{\emph{Eur.
  Phys. J.} {\bfseries C79} (2019) 797}
  [\href{https://arxiv.org/abs/1906.03445}{{\ttfamily 1906.03445}}].

\bibitem{Constantinou:2014rka}
M.~Constantinou, M.~Hadjiantonis and H.~Panagopoulos, \emph{{Renormalization of
  Flavor Singlet and Nonsinglet Fermion Bilinear Operators}},
  \href{https://doi.org/10.22323/1.214.0298}{\emph{PoS} {\bfseries LATTICE2014}
  (2014) 298} [\href{https://arxiv.org/abs/1411.6990}{{\ttfamily 1411.6990}}].

\bibitem{Bali:2016umi}
{\scshape RQCD} collaboration, G.~S. Bali, E.~E. Scholz, J.~Simeth and
  W.~{S\"odner}, \emph{{Lattice simulations with $N_f=2+1$ improved Wilson
  fermions at a fixed strange quark mass}},
  \href{https://doi.org/10.1103/PhysRevD.94.074501}{\emph{Phys. Rev.}
  {\bfseries D94} (2016) 074501}
  [\href{https://arxiv.org/abs/1606.09039}{{\ttfamily 1606.09039}}].

\bibitem{Korcyl:2016ugy}
P.~Korcyl and G.~S. Bali, \emph{{Non-perturbative determination of improvement
  coefficients using coordinate space correlators in $N_f=2+1$ lattice QCD}},
  \href{https://doi.org/10.1103/PhysRevD.95.014505}{\emph{Phys. Rev.}
  {\bfseries D95} (2017) 014505}
  [\href{https://arxiv.org/abs/1607.07090}{{\ttfamily 1607.07090}}].

\bibitem{Luscher:1992zx}
M.~{L\"uscher}, R.~Sommer, U.~Wolff and P.~Weisz, \emph{{Computation of the
  running coupling in the SU(2) Yang-Mills theory}},
  \href{https://doi.org/10.1016/0550-3213(93)90292-W}{\emph{Nucl. Phys.}
  {\bfseries B389} (1993) 247}
  [\href{https://arxiv.org/abs/hep-lat/9207010}{{\ttfamily hep-lat/9207010}}].

\bibitem{Luscher:1993gh}
M.~{L\"uscher}, R.~Sommer, P.~Weisz and U.~Wolff, \emph{{A Precise
  determination of the running coupling in the SU(3) Yang-Mills theory}},
  \href{https://doi.org/10.1016/0550-3213(94)90629-7}{\emph{Nucl. Phys.}
  {\bfseries B413} (1994) 481}
  [\href{https://arxiv.org/abs/hep-lat/9309005}{{\ttfamily hep-lat/9309005}}].

\bibitem{DellaMorte:2004bc}
{\scshape ALPHA} collaboration, M.~Della~Morte, R.~Frezzotti, J.~Heitger,
  J.~Rolf, R.~Sommer and U.~Wolff, \emph{{Computation of the strong coupling in
  QCD with two dynamical flavors}},
  \href{https://doi.org/10.1016/j.nuclphysb.2005.02.013}{\emph{Nucl. Phys.}
  {\bfseries B713} (2005) 378}
  [\href{https://arxiv.org/abs/hep-lat/0411025}{{\ttfamily hep-lat/0411025}}].

\bibitem{Capitani:1998mq}
S.~Capitani, M.~{L\"uscher}, R.~Sommer and H.~Wittig, \emph{{Non-perturbative
  quark mass renormalization in quenched lattice QCD}},
  \href{https://doi.org/10.1016/S0550-3213(00)00163-2,
  10.1016/S0550-3213(98)00857-8}{\emph{Nucl. Phys.} {\bfseries B544} (1999)
  669} [\href{https://arxiv.org/abs/hep-lat/9810063}{{\ttfamily
  hep-lat/9810063}}].

\bibitem{DellaMorte:2005kg}
{\scshape ALPHA} collaboration, M.~Della~Morte, R.~Hoffmann, F.~Knechtli,
  J.~Rolf, R.~Sommer, I.~Wetzorke et~al., \emph{{Non-perturbative quark mass
  renormalization in two-flavor QCD}},
  \href{https://doi.org/10.1016/j.nuclphysb.2005.09.028}{\emph{Nucl. Phys.}
  {\bfseries B729} (2005) 117}
  [\href{https://arxiv.org/abs/hep-lat/0507035}{{\ttfamily hep-lat/0507035}}].

\bibitem{Fritzsch:2013je}
P.~Fritzsch and A.~Ramos, \emph{{The gradient flow coupling in the
  {Schr\"odinger} Functional}},
  \href{https://doi.org/10.1007/JHEP10(2013)008}{\emph{JHEP} {\bfseries 10}
  (2013) 008} [\href{https://arxiv.org/abs/1301.4388}{{\ttfamily 1301.4388}}].

\bibitem{DallaBrida:2016kgh}
{\scshape ALPHA} collaboration, M.~Dalla~Brida, P.~Fritzsch, T.~Korzec,
  A.~Ramos, S.~Sint and R.~Sommer, \emph{{Slow running of the Gradient Flow
  coupling from 200 MeV to 4 GeV in $N_{\rm f}=3$ QCD}},
  \href{https://doi.org/10.1103/PhysRevD.95.014507}{\emph{Phys. Rev.}
  {\bfseries D95} (2017) 014507}
  [\href{https://arxiv.org/abs/1607.06423}{{\ttfamily 1607.06423}}].

\bibitem{Bruno:2017gxd}
{\scshape ALPHA} collaboration, M.~Bruno, M.~Dalla~Brida, P.~Fritzsch,
  T.~Korzec, A.~Ramos, S.~Schaefer et~al., \emph{{QCD Coupling from a
  Nonperturbative Determination of the Three-Flavor $\Lambda$ Parameter}},
  \href{https://doi.org/10.1103/PhysRevLett.119.102001}{\emph{Phys. Rev. Lett.}
  {\bfseries 119} (2017) 102001}
  [\href{https://arxiv.org/abs/1706.03821}{{\ttfamily 1706.03821}}].

\bibitem{Brida:2016flw}
{\scshape ALPHA} collaboration, M.~Dalla~Brida, P.~Fritzsch, T.~Korzec,
  A.~Ramos, S.~Sint and R.~Sommer, \emph{{Determination of the QCD
  $\Lambda$-parameter and the accuracy of perturbation theory at high
  energies}}, \href{https://doi.org/10.1103/PhysRevLett.117.182001}{\emph{Phys.
  Rev. Lett.} {\bfseries 117} (2016) 182001}
  [\href{https://arxiv.org/abs/1604.06193}{{\ttfamily 1604.06193}}].

\bibitem{DallaBrida:2018rfy}
{\scshape ALPHA} collaboration, M.~Dalla~Brida, P.~Fritzsch, T.~Korzec,
  A.~Ramos, S.~Sint and R.~Sommer, \emph{{A non-perturbative exploration of the
  high energy regime in $N_{\mathrm{f}}=3$ QCD}},
  \href{https://doi.org/10.1140/epjc/s10052-018-5838-5}{\emph{Eur. Phys. J.}
  {\bfseries C78} (2018) 372}
  [\href{https://arxiv.org/abs/1803.10230}{{\ttfamily 1803.10230}}].

\bibitem{Luscher:1992an}
M.~{L\"uscher}, R.~Narayanan, P.~Weisz and U.~Wolff, \emph{{The {Schr\"odinger}
  functional: A Renormalizable probe for nonAbelian gauge theories}},
  \href{https://doi.org/10.1016/0550-3213(92)90466-O}{\emph{Nucl. Phys.}
  {\bfseries B384} (1992) 168}
  [\href{https://arxiv.org/abs/hep-lat/9207009}{{\ttfamily hep-lat/9207009}}].

\bibitem{Sint:1993un}
S.~Sint, \emph{{On the {Schr\"odinger} functional in QCD}},
  \href{https://doi.org/10.1016/0550-3213(94)90228-3}{\emph{Nucl. Phys.}
  {\bfseries B421} (1994) 135}
  [\href{https://arxiv.org/abs/hep-lat/9312079}{{\ttfamily hep-lat/9312079}}].

\bibitem{Gasser:1982ap}
J.~Gasser and H.~Leutwyler, \emph{{Quark Masses}},
  \href{https://doi.org/10.1016/0370-1573(82)90035-7}{\emph{Phys. Rept.}
  {\bfseries 87} (1982) 77}.

\bibitem{Sint:1995ch}
S.~Sint and R.~Sommer, \emph{{The Running coupling from the QCD {Schr\"odinger}
  functional: A One loop analysis}},
  \href{https://doi.org/10.1016/0550-3213(96)00020-X}{\emph{Nucl. Phys.}
  {\bfseries B465} (1996) 71}
  [\href{https://arxiv.org/abs/hep-lat/9508012}{{\ttfamily hep-lat/9508012}}].

\bibitem{Sint:1997jx}
S.~Sint and P.~Weisz, \emph{{Further results on O(a) improved lattice QCD to
  one loop order of perturbation theory}},
  \href{https://doi.org/10.1016/S0550-3213(97)00372-6}{\emph{Nucl. Phys.}
  {\bfseries B502} (1997) 251}
  [\href{https://arxiv.org/abs/hep-lat/9704001}{{\ttfamily hep-lat/9704001}}].

\bibitem{Bietenholz:2010jr}
W.~Bietenholz et~al., \emph{{Tuning the strange quark mass in lattice
  simulations}},
  \href{https://doi.org/10.1016/j.physletb.2010.05.067}{\emph{Phys. Lett.}
  {\bfseries B690} (2010) 436}
  [\href{https://arxiv.org/abs/1003.1114}{{\ttfamily 1003.1114}}].

\bibitem{Bietenholz:2011qq}
W.~Bietenholz et~al., \emph{{Flavour blindness and patterns of flavour symmetry
  breaking in lattice simulations of up, down and strange quarks}},
  \href{https://doi.org/10.1103/PhysRevD.84.054509}{\emph{Phys. Rev.}
  {\bfseries D84} (2011) 054509}
  [\href{https://arxiv.org/abs/1102.5300}{{\ttfamily 1102.5300}}].

\bibitem{Bruno:2016avt}
M.~Bruno, \emph{{The energy scale of the 3-flavour $\Lambda$ parameter}}, Ph.D.
  thesis, Humboldt U., Berlin, 2015.
\newblock 10.18452/17516.

\bibitem{Luscher:2012av}
M.~{L\"uscher} and S.~Schaefer, \emph{{Lattice QCD with open boundary
  conditions and twisted-mass reweighting}},
  \href{https://doi.org/10.1016/j.cpc.2012.10.003}{\emph{Comput. Phys. Commun.}
  {\bfseries 184} (2013) 519}
  [\href{https://arxiv.org/abs/1206.2809}{{\ttfamily 1206.2809}}].

\bibitem{Bruno:2014lra}
M.~Bruno, P.~Korcyl, T.~Korzec, S.~Lottini and S.~Schaefer, \emph{{On the
  extraction of spectral quantities with open boundary conditions}},
  \href{https://doi.org/10.22323/1.214.0089}{\emph{PoS} {\bfseries LATTICE2014}
  (2014) 089} [\href{https://arxiv.org/abs/1411.5207}{{\ttfamily 1411.5207}}].

\bibitem{DallaBrida:2018tpn}
M.~Dalla~Brida, T.~Korzec, S.~Sint and P.~Vilaseca, \emph{{High precision
  renormalization of the flavour non-singlet Noether currents in lattice QCD
  with Wilson quarks}},
  \href{https://doi.org/10.1140/epjc/s10052-018-6514-5}{\emph{Eur. Phys. J.}
  {\bfseries C79} (2019) 23}
  [\href{https://arxiv.org/abs/1808.09236}{{\ttfamily 1808.09236}}].

\bibitem{Sint:2010eh}
S.~Sint, \emph{{The Chirally rotated Schr\"odinger functional with Wilson
  fermions and automatic $O(a)$ improvement}},
  \href{https://doi.org/10.1016/j.nuclphysb.2011.02.002}{\emph{Nucl. Phys.}
  {\bfseries B847} (2011) 491}
  [\href{https://arxiv.org/abs/1008.4857}{{\ttfamily 1008.4857}}].

\bibitem{Sint:2010xy}
S.~Sint and B.~Leder, \emph{{Testing universality and automatic $O(a)$
  improvement in massless lattice QCD with Wilson quarks}}, {\emph{PoS}
  {\bfseries LATTICE2010} (2010) 265}
  [\href{https://arxiv.org/abs/1012.2500}{{\ttfamily 1012.2500}}].

\bibitem{Brida:2014zwa}
M.~Dalla~Brida and S.~Sint, \emph{{A dynamical study of the chirally rotated
  {Schr\"odinger} functional in QCD}}, {\emph{PoS} {\bfseries LATTICE2014}
  (2014) 280} [\href{https://arxiv.org/abs/1412.8022}{{\ttfamily 1412.8022}}].

\bibitem{Wolff:2003sm}
{\scshape ALPHA} collaboration, U.~Wolff, \emph{{Monte Carlo errors with less
  errors}}, \href{https://doi.org/10.1016/S0010-4655(03)00467-3,
  10.1016/j.cpc.2006.12.001}{\emph{Comput. Phys. Commun.} {\bfseries 156}
  (2004) 143} [\href{https://arxiv.org/abs/hep-lat/0306017}{{\ttfamily
  hep-lat/0306017}}].

\bibitem{Schaefer:2010hu}
{\scshape ALPHA} collaboration, S.~Schaefer, R.~Sommer and F.~Virotta,
  \emph{{Critical slowing down and error analysis in lattice QCD simulations}},
  \href{https://doi.org/10.1016/j.nuclphysb.2010.11.020}{\emph{Nucl. Phys.}
  {\bfseries B845} (2011) 93}
  [\href{https://arxiv.org/abs/1009.5228}{{\ttfamily 1009.5228}}].

\bibitem{virotta:phdthesis}
F.~Virotta, \emph{Critical slowing down and error analysis of lattice QCD
  simulations}, Ph.D. thesis, Humboldt-Universit{\"a}t zu Berlin,
  Mathematisch-Naturwissenschaftliche Fakult{\"a}t I, 2012.

\bibitem{dobs:alpha}
{\scshape ALPHA} collaboration, H.~Simma, R.~Sommer and F.~Virotta,
  \emph{{General error computation in lattice gauge theory}}, {\emph{ALPHA
  Collaboration internal notes} (2012-2014) }.

\bibitem{Ramos:2018vgu}
A.~Ramos, \emph{{Automatic differentiation for error analysis of Monte Carlo
  data}}, \href{https://doi.org/10.1016/j.cpc.2018.12.020}{\emph{Comput. Phys.
  Commun.} {\bfseries 238} (2019) 19}
  [\href{https://arxiv.org/abs/1809.01289}{{\ttfamily 1809.01289}}].

\bibitem{Colangelo:2005gd}
G.~Colangelo, S.~{D\"urr} and C.~Haefeli, \emph{{Finite volume effects for
  meson masses and decay constants}},
  \href{https://doi.org/10.1016/j.nuclphysb.2005.05.015}{\emph{Nucl. Phys.}
  {\bfseries B721} (2005) 136}
  [\href{https://arxiv.org/abs/hep-lat/0503014}{{\ttfamily hep-lat/0503014}}].

\bibitem{Allton:2008pn}
{\scshape RBC-UKQCD} collaboration, C.~Allton et~al., \emph{{Physical Results
  from 2+1 Flavor Domain Wall QCD and SU(2) Chiral Perturbation Theory}},
  \href{https://doi.org/10.1103/PhysRevD.78.114509}{\emph{Phys. Rev.}
  {\bfseries D78} (2008) 114509}
  [\href{https://arxiv.org/abs/0804.0473}{{\ttfamily 0804.0473}}].

\bibitem{Blum:2014tka}
{\scshape RBC, UKQCD} collaboration, T.~Blum et~al., \emph{{Domain wall QCD
  with physical quark masses}},
  \href{https://doi.org/10.1103/PhysRevD.93.074505}{\emph{Phys. Rev.}
  {\bfseries D93} (2016) 074505}
  [\href{https://arxiv.org/abs/1411.7017}{{\ttfamily 1411.7017}}].

\bibitem{Durr:2010vn}
S.~{D\"urr}, Z.~Fodor, C.~Hoelbling, S.~D. Katz, S.~Krieg, T.~Kurth et~al.,
  \emph{{Lattice QCD at the physical point: light quark masses}},
  \href{https://doi.org/10.1016/j.physletb.2011.05.053}{\emph{Phys. Lett.}
  {\bfseries B701} (2011) 265}
  [\href{https://arxiv.org/abs/1011.2403}{{\ttfamily 1011.2403}}].

\bibitem{Durr:2010aw}
S.~{D\"urr}, Z.~Fodor, C.~Hoelbling, S.~Katz, S.~Krieg et~al., \emph{{Lattice
  QCD at the physical point: Simulation and analysis details}},
  \href{https://doi.org/10.1007/JHEP08(2011)148}{\emph{JHEP} {\bfseries 1108}
  (2011) 148} [\href{https://arxiv.org/abs/1011.2711}{{\ttfamily 1011.2711}}].

\bibitem{McNeile:2010ji}
C.~McNeile, C.~T.~H. Davies, E.~Follana, K.~Hornbostel and G.~P. Lepage,
  \emph{{High-Precision c and b Masses, and QCD Coupling from Current-Current
  Correlators in Lattice and Continuum QCD}},
  \href{https://doi.org/10.1103/PhysRevD.82.034512}{\emph{Phys. Rev.}
  {\bfseries D82} (2010) 034512}
  [\href{https://arxiv.org/abs/1004.4285}{{\ttfamily 1004.4285}}].

\bibitem{Maezawa:2016vgv}
Y.~Maezawa and P.~Petreczky, \emph{{Quark masses and strong coupling constant
  in 2+1 flavor QCD}},
  \href{https://doi.org/10.1103/PhysRevD.94.034507}{\emph{Phys. Rev.}
  {\bfseries D94} (2016) 034507}
  [\href{https://arxiv.org/abs/1606.08798}{{\ttfamily 1606.08798}}].

\bibitem{Bazavov:2009fk}
{\scshape MILC} collaboration, A.~Bazavov et~al., \emph{{MILC results for light
  pseudoscalars}}, \href{https://doi.org/10.22323/1.086.0007}{\emph{PoS}
  {\bfseries CD09} (2009) 007}
  [\href{https://arxiv.org/abs/0910.2966}{{\ttfamily 0910.2966}}].

\bibitem{Bazavov:2010yq}
A.~Bazavov et~al., \emph{{Staggered chiral perturbation theory in the
  two-flavor case and SU(2) analysis of the MILC data}}, {\emph{PoS} {\bfseries
  LATTICE2010} (2010) 083} [\href{https://arxiv.org/abs/1011.1792}{{\ttfamily
  1011.1792}}].

\bibitem{Rupak:2002sm}
G.~Rupak and N.~Shoresh, \emph{{Chiral perturbation theory for the Wilson
  lattice action}},
  \href{https://doi.org/10.1103/PhysRevD.66.054503}{\emph{Phys. Rev.}
  {\bfseries D66} (2002) 054503}
  [\href{https://arxiv.org/abs/hep-lat/0201019}{{\ttfamily hep-lat/0201019}}].

\bibitem{Bar:2003mh}
O.~{B\"ar}, G.~Rupak and N.~Shoresh, \emph{{Chiral perturbation theory at
  $O(a^2)$ for lattice QCD}},
  \href{https://doi.org/10.1103/PhysRevD.70.034508}{\emph{Phys. Rev.}
  {\bfseries D70} (2004) 034508}
  [\href{https://arxiv.org/abs/hep-lat/0306021}{{\ttfamily hep-lat/0306021}}].

\end{thebibliography}\endgroup

\end{document}